\documentclass[aps, prl, showpacs,twocolumn,letterpaper,superscriptaddress,longbibliography,floatfix]{revtex4-1}
\usepackage{graphicx}
\usepackage{amsmath}
\usepackage{latexsym}
\usepackage{amsfonts}
\usepackage{amssymb}
\usepackage{array}
\usepackage{color}
\usepackage{verbatim}
\usepackage{braket}
\usepackage{mathtools}
\graphicspath{{./figures/}}

\begin{document}
\title{Phase Locking between Two All-Optical Quantum Memories}
\author{Fumiya~Okamoto}
\affiliation{Department of Applied Physics, School of Engineering, The University of Tokyo, 7-3-1 Hongo, Bunkyo-ku, Tokyo 113-8656, Japan}
\author{Mamoru~Endo}
\email{endo@ap.t.u-tokyo.ac.jp}
\affiliation{Department of Applied Physics, School of Engineering, The University of Tokyo, 7-3-1 Hongo, Bunkyo-ku, Tokyo 113-8656, Japan}
\author{Mikihisa~Matsuyama}
\affiliation{Department of Applied Physics, School of Engineering, The University of Tokyo, 7-3-1 Hongo, Bunkyo-ku, Tokyo 113-8656, Japan}
\author{Yuya~Ishizuka}
\affiliation{Department of Applied Physics, School of Engineering, The University of Tokyo, 7-3-1 Hongo, Bunkyo-ku, Tokyo 113-8656, Japan}
\author{Yang~Liu}
\affiliation{State Key Laboratory of Quantum Optics and Quantum Optics Devices, Collaborative Innovation Center of Extreme Optics, Institute of Opto-Electronics, Shanxi University, Taiyuan 030006, China}
\author{Rei~Sakakibara}
\affiliation{Department of Applied Physics, School of Engineering, The University of Tokyo, 7-3-1 Hongo, Bunkyo-ku, Tokyo 113-8656, Japan}
\author{Yosuke~Hashimoto}
\affiliation{Department of Applied Physics, School of Engineering, The University of Tokyo, 7-3-1 Hongo, Bunkyo-ku, Tokyo 113-8656, Japan}
\author{Jun-ichi~Yoshikawa}
\email{yoshikawa@ap.t.u-tokyo.ac.jp}
\affiliation{Department of Applied Physics, School of Engineering, The University of Tokyo, 7-3-1 Hongo, Bunkyo-ku, Tokyo 113-8656, Japan}
\author{Peter~van~Loock}
\affiliation{Institute of Physics, Johannes Gutenberg-Universit\"{a}t Mainz, Staudingerweg 7,
55099 Mainz, Germany}
\author{Akira~Furusawa}
\email{akiraf@ap.t.u-tokyo.ac.jp}
\affiliation{Department of Applied Physics, School of Engineering, The University of Tokyo, 7-3-1 Hongo, Bunkyo-ku, Tokyo 113-8656, Japan}

\date{\today}

\begin{abstract}
Optical approaches to quantum computation require the creation of multi-mode photonic quantum states in a controlled fashion. 
Here we experimentally demonstrate phase locking of two all-optical quantum memories, based on a concatenated cavity system with phase reference beams, for the time-controlled release of two-mode entangled single-photon states. The release time for each mode can be independently determined. The generated states are characterized by two-mode optical homodyne tomography. Entanglement and nonclassicality are preserved for release-time differences up to 400 ns, confirmed by logarithmic negativities and Wigner-function negativities, respectively.
\end{abstract}

\maketitle
%%% Introduction %%%%%
Optical quantum information processing based on continuous variables, where quantum information is encoded in traveling electromagnetic field modes, is one of the most promising approaches to efficient, practical quantum computation \cite{Lloyd1999}.
In particular, the generation of scalable continuous-variable cluster states (CVCSs)---highly entangled multi-mode Gaussian states---was demonstrated recently in a time-domain multiplexing scheme~\cite{Yokoyama2013,Yoshikawa2016,Asavanant2019,Larsen2019}.
Moreover, an experiment was recently reported, in which 100 Gaussian gate operations were executed via such a temporal resource state at a high clock rate \cite{Asavanant2020}.
The remaining challenge towards universal fault-tolerant quantum computation including quantum error correction is to incorporate non-Gaussian states into the time-multiplexed CVCS.
A promising implementation for fault-tolerant universal quantum computation has been proposed \cite{Alexander2018}, where non-Gaussian quantum states, such as cubic-phase states \cite{Miyata2016} and Gottesman-Kitaev-Preskill states \cite{Gottesman2001}, are injected into the time-multiplexed CVCS.
However, the bottleneck here is that currently such non-Gaussian states can be only probabilistically generated in a heralding fashion. As a possible remedy, optical quantum-state sources with phase-locked quantum memories can be employed to temporarily store and eventually release the optical wave packets (WPs) at appropriate times ($t_1,t_2,\dots$), as illustrated in Fig.~\ref{Fig_CVCS}.
\begin{figure}
\centering
\includegraphics{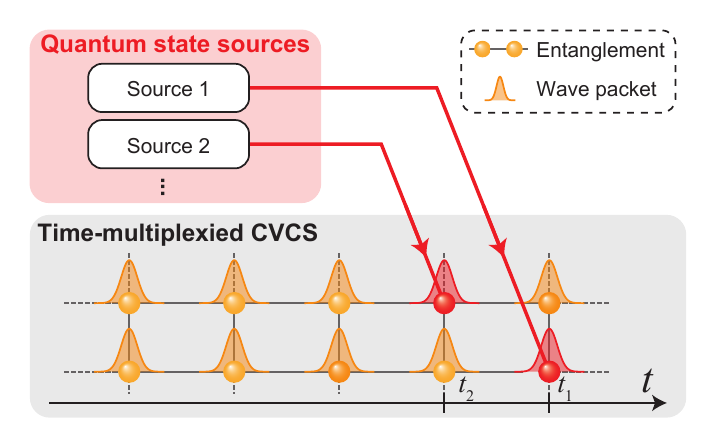}
\caption{Possible concept for fault-tolerant universal quantum computation based on a time-multiplexed CVCS, as proposed in Ref.~\cite{Alexander2018}.
Wave packets (WPs) of quantum states of light, which are generated by phase-locked non-classical, non-Gaussian state sources, are injected into the CVCS in a controlled fashion at certain times $t=t_1, t_2,\dots$.
}
\label{Fig_CVCS}
\end{figure}
\begin{figure*}
\centering
\includegraphics[width = 0.9\textwidth,clip]{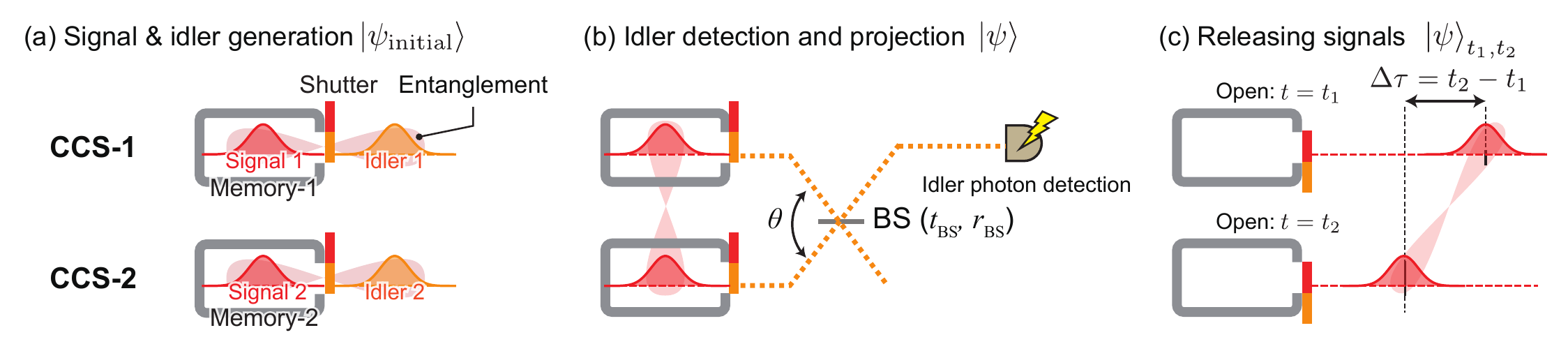}
\caption{Conceptual diagram of release timing control by two quantum memories.
(a) Generation of entanglement between signal (red) and idler (orange) modes in each memory cavity (Eq.~\eqref{eq:two-mode-squeezer}).
Only the idler modes are released.
(b) Idler detection and projection (Eq.~\eqref{eq:dual-rail}).
Arbitrary entangled states can be generated by tuning the beam splitting ratio ($t_\text{BS},r_\text{BS}$) and the phase between two phase references from each memory ($\theta$).
(c) Quasi-on-demand release of the entangled state by opening the shutters for the signal modes independently ($t_1, t_2$).}
\label{Fig_scheme}
\end{figure*}

Although storage of quantum states of light has been demonstrated in various ways \cite{Lvovsky2009, Lai2018, Clausen2011}, only in recent years nonclassical optical states were kept in a quantum memory to an extent that allowed to preserve the negative values of the states' Wigner functions~\cite{Bimbard2014,Bouillard2019,Yoshikawa2014,Hashimoto2019}.
Here the negativity of the Wigner function determined by homodyne tomography indicates strong nonclassicality, which, however, is easily degraded by loss or phase fluctuations.
Most notably, a novel all-optical concatenated cavity system (CCS) composed of a memory and a shutter cavity may serve as an elementary tool especially for CVCS-based universal quantum computation, because of the system's low-loss configuration for various wavelength regimes.
A single CCS was shown to store a nonclassical single-photon state preserving its Wigner-function negativity \cite{Yoshikawa2014}. Subsequently, two CCSs were utilized in order to synchronize the release of two single photons and to create an entangled Hong-Ou-Mandel state verified by full homodyne tomography \cite{Makino2016}.
More recently, the CCS was extended to store and release a non-Gaussian, phase-sensitive, single-mode superposition state by introducing phase probe beams into the CCS~\cite{Hashimoto2019}.
The ultimate next step towards applications such as CVCS quantum computing is now to synchronize multiple phase-sensitive memories by introducing a phase-locking mechanism between them and to release the stored WPs independently in a time-controlled fashion.

Here, we demonstrate phase locking between two timing controllable phase-sensitive memories. This enables us to release, in a controlled fashion, single-photon two-mode entangled states created inside the memories by a heralding mechanism. 
More specifically, we create quasi on demand optical quantum states such as $\alpha\ket{0}_1\ket{1}_2+\beta \text{e}^{\text{i}\theta}\ket{1}_1\ket{0}_2$, where $\alpha, \beta, \theta \in \mathbb{R}$, $\ket{n}$ represents an $n$-photon state (here $n=0,1$), and the subscripts $_{1,2}$ denote mode numbers.
The system can tailor the parameters ($\alpha, \beta, \theta$), in principle, arbitrarily and it can independently control the release times ($t_1, t_2$) for the two optical modes.
Here in the main text, we focus on the simplest, symmetric $\theta=0$ case, $(\alpha,\beta,\theta)=(1/\sqrt{2}, 1/\sqrt{2}, 0)$, for which we confirmed that the two modes maintain their quantum coherence and nonclassicality during storage times of up to 400 ns.
Some additional experimental results with a different choice of parameters for the released two-mode quantum states such as $(1/\sqrt{3},\sqrt{2/3}, \pi), (1/\sqrt{2},1/\sqrt{2}, \pi)$ and $(1/\sqrt{2},1/\sqrt{2}, 5\pi/6)$, including different emission timings, are presented in the Supplemental Material \cite{Supplemental}.
The phase-controlled synchronization among multi-mode quantum systems as demonstrated here can be directly utilized for fault-tolerant large-scale quantum computation with CVCSs as proposed in Ref.~\cite{Alexander2018}.
In our experiment, the two-mode states have fixed photon number one,
similar to a one-photon two-mode qubit state (so-called dual-rail qubit).
In other words, our experiment is probably the first demonstration for a quasi-on-demand
generation of an arbitrary dual-rail qubit (experimentally drawn from a finite representative set of qubit states).
However, we stress that this particular manifestation of our work highlights only a narrow aspect of
our system, which can be more generally applied to higher photon and mode numbers as required in CVCS quantum computation. As for higher photon numbers, it is already theoretically known that we can directly generate arbitrary two-mode states with a fixed photon number, $\sum_{n=0}^{N}c_{n,N-n}\ket{n}_1\ket{N-n}_2$ where $c_{k,l}\in\mathbb{C}$ \cite{Yoshikawa2018}. This corresponds to encoding a single spin of arbitrary size into two optical field modes \cite{JulianSeymourSchwinger}.
The resulting larger Hilbert spaces may then also contain bosonic error correction codes,
for example, $\alpha(\ket{4}_1\ket{0}_2+\ket{0}_1\ket{4}_2)/\sqrt{2}+\beta\ket{2}_1\ket{2}_2$
adapted for protection against photon loss \cite{Chuang1997}. By adding more modes, N00N states for quantum-enhanced metrology \cite{Jin2013c,Muller2017} or again for quantum error correction \cite{Bergmann2016} may be producible in a controlled way.
To our knowledge, this is the first demonstration of phase locking multi-mode optical quantum systems with time-controlled memories, verified by two-mode optical homodyne tomography.
\begin{figure*}
\centering
\includegraphics[width = 0.9\textwidth,clip]{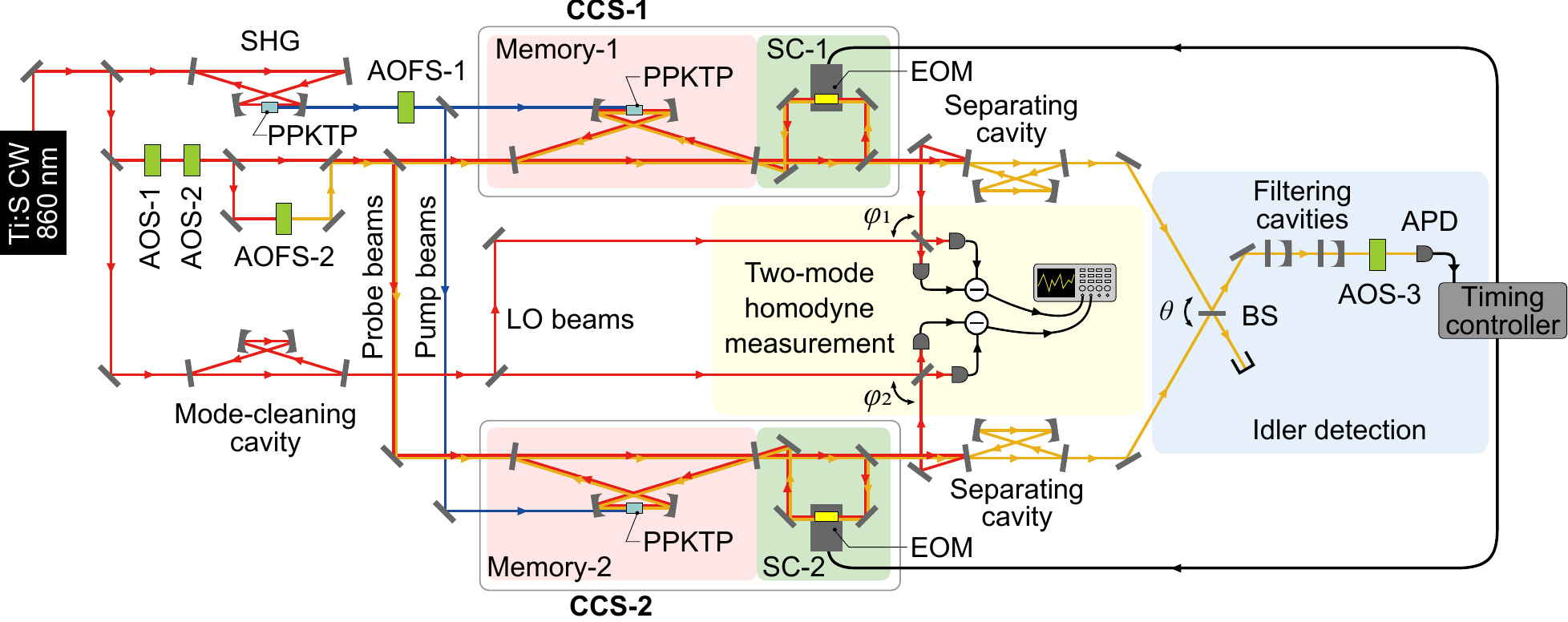}
\caption{Experimental apparatus.
The light source is a 860 nm continuous-wave (CW) Ti:Sapphire laser.
Red and orange lines stand for the signal and idler beams, respectively.
Blue lines represent the pump beams.
Black lines stand for electric signals.
LO: local oscillator, CCS: concatenated cavity system, Memory: memory cavity based on NOPO with a cavity length of 1.4m, SC: shutter cavity with a length of 0.7m, PPKTP: periodically-poled KTiOPO$_4$ crystal (type-0), EOM: electro-optic modulator, AOFSs: acousto-optic frequency shifters, AOSs: acousto-optic switches, BS: beam splitter for setting superposition parameters $(\alpha, \beta)$, APD: silicon avalanche photodiode.}
\label{Fig_Setup}
\end{figure*}

%%% Heralded generation of dual rail qubit %%%%%
Let us explain how to generate the two-mode quantum states $\alpha\ket{0}_1\ket{1}_2+\beta \text{e}^{\text{i}\theta}\ket{1}_1\ket{0}_2$ by means of two non-degenerate optical parametric oscillators (NOPOs), linear optics, and single-photon detection.
The (unnormalized) initial four-mode state $\ket{\psi_\text{initial}}$ generated from the two NOPOs can be written as
\begin{align}
    \ket{\psi_\text{initial}}=\sum_{n_1=0}^\infty q_1^{n_1} \ket{n_1}_{s_1} \ket{n_1}_{i_1}\otimes\sum_{n_2=0}^\infty q_2^{n_2}\ket{n_2}_{s_2}\ket{n_2}_{i_2} \label{eq:two-mode-squeezer},
\end{align}
where $s,i$ represent ``signal'' and ``idler'' modes, and $q_{1,2} $ include the pump amplitudes for the NOPOs.
The idler modes from the two NOPOs are combined at a beam splitter (BS).
When a single photon is detected at one of the outputs of the BS, the signal fields are projected onto a two-mode entangled state $\ket{\psi}$,
\begin{align}
   &_{i_1}\bra{0}_{i_2}\bra{1}\hat{U}_{i_1,i_2}(t_\text{BS},r_\text{BS})\ket{\psi_{\text{initial}}}\nonumber\\
   &=t_\text{BS}q_2\ket{0}_{s_1}\ket{1}_{s_2}+r_\text{BS}q_1\ket{1}_{s_1}\ket{0}_{s_2} \nonumber\\
    &\propto \alpha\ket{0}_{s_1}\ket{1}_{s_2}+\beta \text{e}^{\text{i}\theta}\ket{1}_{s_1}\ket{0}_{s_2}=\ket{\psi}\label{eq:dual-rail},
\end{align}
where $\hat{U}_{i_1,i_2}(t_\text{BS},r_\text{BS})$ is a BS operator acting on modes $i_1$ and $i_2$.
The parameters $t_\text{BS}, r_\text{BS}\in \mathbb{C}$  are transmission and reflection coefficients of the BS, respectively, satisfying $|t_\text{BS}|^2+|r_\text{BS}|^2=1$.
Finally, the parameters $(\alpha, \beta, \theta)$ in Eq.~\eqref{eq:dual-rail} are determined by $t_\text{BS},r_\text{BS}$ and $q_{1,2}$.
Note that the state projection via single-photon detection at one output port of the BS is expressed by $_{i_1}\bra{0}_{i_2}\bra{1}\hat{U}_{i_1,i_2}(t_\text{BS},r_\text{BS})$ under a weak-pumping condition ($|q_{1,2}|\ll 1$), while
the detector employed in the experiment was an avalanche photodiode (APD).
This method can be extended to the generation of two-mode states with higher photon numbers \cite{Yoshikawa2018}.
Note that, in the following, we omit the subscripts for simplicity.

Figure~\ref{Fig_scheme} shows how to combine the above heralding method with the CCSs and how to control the release times of the two-mode-state WPs.
Non-degenerate parametric down-conversion occurs inside each memory cavity (Memory-1, 2).
The shutters in the CCSs are concatenated shutter cavities (SCs) transmitting either the signal or the idler for suitably tuned resonance frequencies.
At the beginning, the SCs are only open for the idlers, thus the signals are stored in each memory (Fig.~\ref{Fig_scheme}~(a)).
The emitted idlers are combined at the BS with phase $\theta$.
When an idler photon is detected, the resulting state $\ket{\psi}$ emerges in the memory cavities (Fig.~\ref{Fig_scheme}~(b)).
Each shutter is opened after appropriate waiting times $(t_1, t_2)$ by shifting each SC's resonance frequency correspondingly; then the signal WPs are released independently (Fig.~\ref{Fig_scheme}~(c)).
Because the two memories are phase locked via the idler probe beams, the output two-mode 
state can keep its phase relation, although the releases
of the two WPs will typically happen at different times.

The experimental apparatus is illustrated in Fig.~\ref{Fig_Setup} and further details can be found in the Supplemental Material \cite{Supplemental}.
We select the signal and the idler fields to be separated in frequency by the free spectral range (FSR) of the memory cavities.
In order to separate the signal and the idler fields, which are colinearly emitted from the same output coupler of each SC, frequency separating cavities are employed.
Then, the idler modes are combined at a beam splitter (BS) for setting the superposition parameters $(\alpha, \beta, \theta)$.
One of the BS output fields is guided towards an APD through cascaded filtering cavities to eliminate unwanted light.
The probe beams for the idler and the signal are injected into the memory cavities, while the optical frequency of the idler probe is up-shifted with the corresponding acousto-optic frequency shifter (AOFS-2) by 214 MHz, which corresponds to the FSR of the memory cavity.
To probe the phase of the signal and the idler fields in the CCSs, the phase relation among the pump, the signal probe and the idler probe beams are locked via phase-sensitive parametric amplification.
These probe beams are used for phase-locking the idler beams at the BS, and for the
signal-homodyne measurements.
The probe beams are essential for the phase-sensitive memories, but they should be absent during storage and release of the signal quantum states, because strong probe beams would disturb the heralding and homodyne signals.
Optical switches based on acousto-optic modulators (AOS-1, 2) with periods of 200 $\mu$s are used as a beam chopper.
%Optical switches based on acousto-optic modulators (AOS-1, 2) are used as a beam chopper with a period of 200 $\mu$s, which is sufficiently longer than the decay time of the probe beams in the cavities \cite{Hashimoto2019}.
As long as the probe beams are on, the system locks the cavities and the phase relations.
Storage and release of the quantum states as well as homodyne measurements are done when the probe beams are off.

\begin{figure*}
\centering
\includegraphics[width = 1\textwidth, clip]{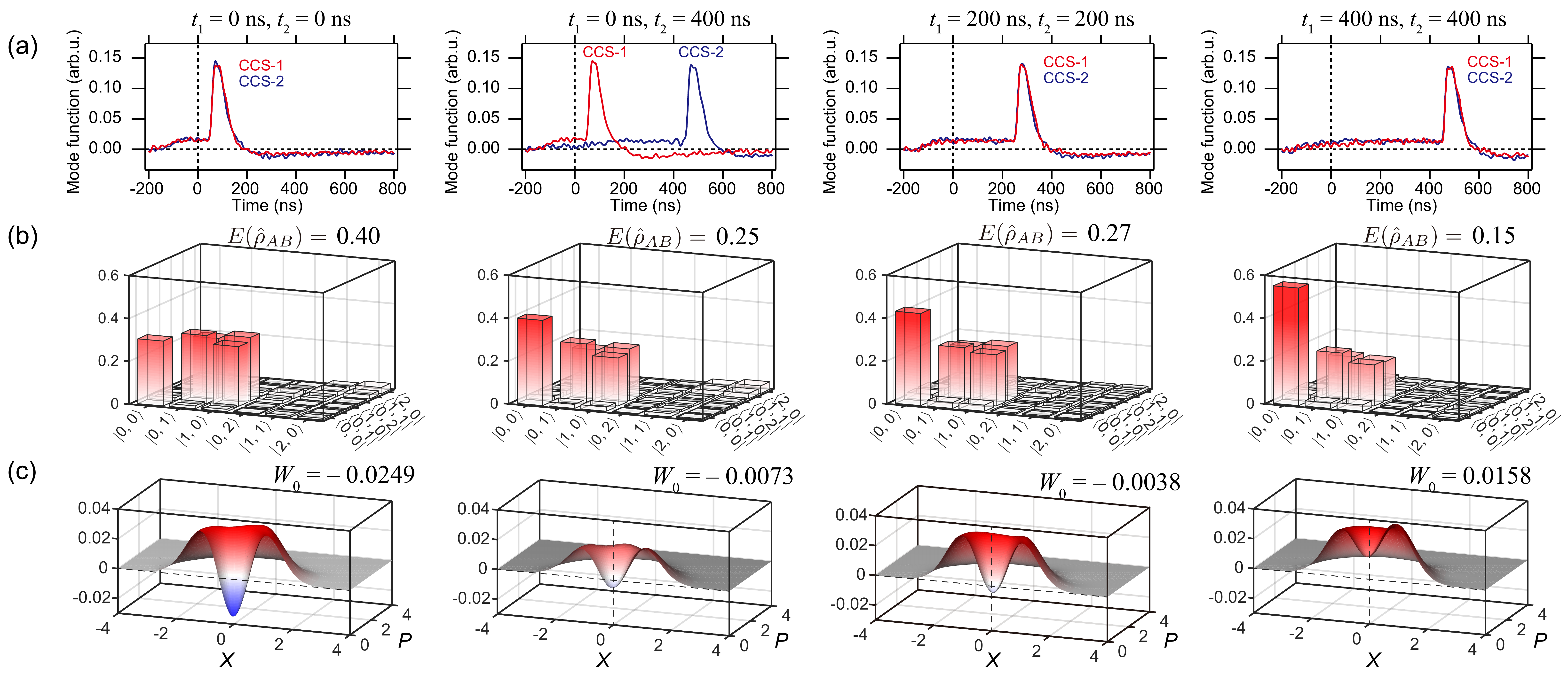}
\caption{Experimental results. (a) Experimentally determined envelopes of the released WP amplitudes from CCS-1 (red traces) and CCS-2 (blue traces).
The idler-photon detection event corresponds to 0ns.
(b) Absolute values of density matrix elements with experimentally determined logarithmic negativities.
(c) Calculated cross sections of the Wigner function $W(X, P, X, P)$ with the values at the origin.
The emission times $(t_1, t_2)$ are $(0\text{ns}, 0\text{ns}), (0\text{ns}, 400\text{ns})$, $(200\text{ns}, 200\text{ns})$, and $(400\text{ns}, 400\text{ns})$ from the left to the right columns. }
\label{Fig_result}
\end{figure*}
In the main text, we refer to a BS with a reflectivity of 50\% and a phase $\theta = 0$.
In the Supplemental Material \cite{Supplemental}, we show results with other BSs and we lock to various phases corresponding to different two-mode quantum states.
In order for each memory's counts to be the same, the pump powers into the memory cavities are suitably adjusted (approximately 2 mW each).
The average count rate of each CCS was set to 200 counts per second (cps) and the fake count rate by stray light is at a 10 cps level.
After receiving the heralding signal and predetermined waiting times ($t_1,t_2$), a timing controller based on a field-programmable gate array sends trigger signals to open the SCs.
Each shutter is opened by applying a high voltage (900 V) to an electro-optic modulator (EOM).
The released signal fields are measured by two optical homodyne detectors.
Note that the resonance frequency of the memory cavities is detuned (300 kHz) from the frequency of the local oscillator (LO) beams to suppress the occurrence of unwanted photons stored by the memory, which is caused by the scattering of the strong LO beams.
This detuning leads to a rotation of the released quantum states for $t_1\ne t_2$ \cite{Supplemental}.

We performed balanced optical homodyne measurements for the signal fields around the heralding signals at various LOs' phases $\varphi_1, \varphi_2$, where 3000 pairs of waveforms for each of 49 homodyne bases ($\varphi_1, \varphi_2$) were acquired by an oscilloscope.
The LOs' phases were stabilized via the heterodyne beat notes between the signal probe and the LO beams.

%%%%%%%%%%%%%%%% Experimental results%%%%%%%%%%%%%%%%%%%%%%
In a preliminary experiment, the envelopes of the signal WPs for the single-photon states were determined by homodyne measurements \cite{Macrae2012,Morin2013,Yoshikawa2014}.
The envelopes of the released WP amplitudes for different emission times are shown in Fig.~\ref{Fig_result}~(a).
The emission times are correctly shifted and the shape of the envelopes is independent of the storage times.
The idler-photon detection event corresponds to 0 ns (Fig.~\ref{Fig_scheme} (b)).
The occurring gap of about 40 ns between $t=0$ and the rising edges of the envelopes when $t_1=t_2=0$ is due to a latency of the electronics.
Any non-zero positive values before the photon emissions correspond to leakage of the memories, but the contribution of this leakage can be ignored \cite{Yoshikawa2014}.

As a final result, we stored and released two-mode entangled states to infer their density matrices.
The density matrices for the WPs in Fig.~\ref{Fig_result} (a) are shown in Fig.~\ref{Fig_result} (b).
The  off-diagonal terms, $\ket{0,1}\bra{1,0}$ and $\ket{1,0}\bra{0,1}$, indicate the presence of entanglement.
The diagonal and the off-diagonal terms decrease at the same rate.
This means that the system preserves the phase information even though small intracavity losses degrade the quantum states during their storage.
Figures~\ref{Fig_result}~(a) and (b) give evidence that the CCSs are indeed sufficiently phase locked and that the memories release entangled states in the various target WPs preserving their phase relations.
The  storage times are 1.42 $\mu$s (CCS-1) and 1.29 $\mu$s (CCS-2).
The logarithmic negativities $E(\hat{\rho}_{AB}) = \log_2||\hat{\rho}_{AB}^{T_B}||$ are calculated and presented in Fig.~\ref{Fig_result} (b) to confirm the entanglement \cite{Plenio2005,Vidal2002,Plenio1998}, where $||\cdot||$ and $\cdot^{T_B}$ represent the trace norm and partial transpose, respectively.
Here, the calculation of the logarithmic negativity is done in the re-normalized subspace spanned by $\{\ket{0,0}, \ket{0,1}, \ket{1,0}, \ket{1,1}\}$ \cite{Supplemental}.
A non-zero value of the logarithmic negativity indicates that the state is entangled, which is still observable in our experiment even when $t_1=t_2=400$ ns.
Two-mode Wigner functions $W(x_1, p_1, x_2, p_2)$ are calculated from the density matrices to confirm that our system is able to preserve strong nonclassicality.
The cross sections $W(X, P, X, P)$ and the values at the phase-space origin ($W_0$) are shown in Fig.~\ref{Fig_result} (c).
The negative values around the origin indicate that our CCSs can store and release fragile, nonclassical states for emission times $(0\ \text{ns}, 0\ \text{ns}), (0\ \text{ns}, 400\ \text{ns}), (200\ \text{ns}, 200\ \text{ns})$.
Even though the negativities of the Wigner function can be easily degraded, our system successfully preserves them thanks to its low-loss configuration.

%%%%%%%%%%%% conclusion
In conclusion, we experimentally demonstrated the coherent synchronization and hence the quasi-on-demand release of two-mode quantum states of light by means of two time-controlled, all-optical, phase-locked memories.
Via full control over the phase relation of the two memories, the release times of the signal modes can be independently adjusted to keep the entanglement and the nonclassicality of the optical quantum states.
We experimentally prepared a representative set of single-photon dual-rail qubit states quasi on demand. Our scheme is also directly applicable to the controlled creation of optical states with more modes and more photons including logical qubits of bosonic quantum error corection codes
with a fixed photon number sufficiently exceeding one. Since the exploitation of optical quantum states with non-zero photon-number variance depends in particular on the reliable manipulation of their relative phases, our scheme also provides a fundamental solution for this broader class of applications.
Therefore, ultimately, it is applicable to optical measurement-based, fault-tolerant universal quantum information processing.

This work was partly supported by JSPS KAKENHI (Grants No. 18H01149 and 18H05207), CREST (Grant No. JPMJCR15N5) of JST, UTokyo Foundation, and donations from Nichia Corporation of Japan.
F.O. and Y.H. acknowledge support from ALPS. M.M. acknowledges support from FoPM. P.v.L. acknowledges QLinkX (BMBF, Germany) and ShoQC (Quantera, EU/BMBF).
\bibliographystyle{apsrev4-1}
\bibliography{myPapers-20201006_twomode}

\begin{thebibliography}{29}%
\makeatletter
\providecommand \@ifxundefined [1]{%
 \@ifx{#1\undefined}
}%
\providecommand \@ifnum [1]{%
 \ifnum #1\expandafter \@firstoftwo
 \else \expandafter \@secondoftwo
 \fi
}%
\providecommand \@ifx [1]{%
 \ifx #1\expandafter \@firstoftwo
 \else \expandafter \@secondoftwo
 \fi
}%
\providecommand \natexlab [1]{#1}%
\providecommand \enquote  [1]{``#1''}%
\providecommand \bibnamefont  [1]{#1}%
\providecommand \bibfnamefont [1]{#1}%
\providecommand \citenamefont [1]{#1}%
\providecommand \href@noop [0]{\@secondoftwo}%
\providecommand \href [0]{\begingroup \@sanitize@url \@href}%
\providecommand \@href[1]{\@@startlink{#1}\@@href}%
\providecommand \@@href[1]{\endgroup#1\@@endlink}%
\providecommand \@sanitize@url [0]{\catcode `\\12\catcode `\$12\catcode
  `\&12\catcode `\#12\catcode `\^12\catcode `\_12\catcode `\%12\relax}%
\providecommand \@@startlink[1]{}%
\providecommand \@@endlink[0]{}%
\providecommand \url  [0]{\begingroup\@sanitize@url \@url }%
\providecommand \@url [1]{\endgroup\@href {#1}{\urlprefix }}%
\providecommand \urlprefix  [0]{URL }%
\providecommand \Eprint [0]{\href }%
\providecommand \doibase [0]{http://dx.doi.org/}%
\providecommand \selectlanguage [0]{\@gobble}%
\providecommand \bibinfo  [0]{\@secondoftwo}%
\providecommand \bibfield  [0]{\@secondoftwo}%
\providecommand \translation [1]{[#1]}%
\providecommand \BibitemOpen [0]{}%
\providecommand \bibitemStop [0]{}%
\providecommand \bibitemNoStop [0]{.\EOS\space}%
\providecommand \EOS [0]{\spacefactor3000\relax}%
\providecommand \BibitemShut  [1]{\csname bibitem#1\endcsname}%
\let\auto@bib@innerbib\@empty
%</preamble>
\bibitem [{\citenamefont {Lloyd}\ and\ \citenamefont
  {Braunstein}(1999)}]{Lloyd1999}%
  \BibitemOpen
  \bibfield  {author} {\bibinfo {author} {\bibfnamefont {S.}~\bibnamefont
  {Lloyd}}\ and\ \bibinfo {author} {\bibfnamefont {S.~L.}\ \bibnamefont
  {Braunstein}},\ }\href {\doibase 10.1103/PhysRevLett.82.1784} {\bibfield
  {journal} {\bibinfo  {journal} {Physical Review Letters}\ }\textbf {\bibinfo
  {volume} {82}},\ \bibinfo {pages} {1784} (\bibinfo {year}
  {1999})}\BibitemShut {NoStop}%
\bibitem [{\citenamefont {Yokoyama}\ \emph {et~al.}(2013)\citenamefont
  {Yokoyama}, \citenamefont {Ukai}, \citenamefont {Armstrong}, \citenamefont
  {Sornphiphatphong}, \citenamefont {Kaji}, \citenamefont {Suzuki},
  \citenamefont {Yoshikawa}, \citenamefont {Yonezawa}, \citenamefont
  {Menicucci},\ and\ \citenamefont {Furusawa}}]{Yokoyama2013}%
  \BibitemOpen
  \bibfield  {author} {\bibinfo {author} {\bibfnamefont {S.}~\bibnamefont
  {Yokoyama}}, \bibinfo {author} {\bibfnamefont {R.}~\bibnamefont {Ukai}},
  \bibinfo {author} {\bibfnamefont {S.~C.}\ \bibnamefont {Armstrong}}, \bibinfo
  {author} {\bibfnamefont {C.}~\bibnamefont {Sornphiphatphong}}, \bibinfo
  {author} {\bibfnamefont {T.}~\bibnamefont {Kaji}}, \bibinfo {author}
  {\bibfnamefont {S.}~\bibnamefont {Suzuki}}, \bibinfo {author} {\bibfnamefont
  {J.}~\bibnamefont {Yoshikawa}}, \bibinfo {author} {\bibfnamefont
  {H.}~\bibnamefont {Yonezawa}}, \bibinfo {author} {\bibfnamefont {N.~C.}\
  \bibnamefont {Menicucci}}, \ and\ \bibinfo {author} {\bibfnamefont
  {A.}~\bibnamefont {Furusawa}},\ }\href {\doibase 10.1038/nphoton.2013.287}
  {\bibfield  {journal} {\bibinfo  {journal} {Nature Photonics}\ }\textbf
  {\bibinfo {volume} {7}},\ \bibinfo {pages} {982} (\bibinfo {year}
  {2013})}\BibitemShut {NoStop}%
\bibitem [{\citenamefont {Yoshikawa}\ \emph {et~al.}(2016)\citenamefont
  {Yoshikawa}, \citenamefont {Yokoyama}, \citenamefont {Kaji}, \citenamefont
  {Sornphiphatphong}, \citenamefont {Shiozawa}, \citenamefont {Makino},\ and\
  \citenamefont {Furusawa}}]{Yoshikawa2016}%
  \BibitemOpen
  \bibfield  {author} {\bibinfo {author} {\bibfnamefont {J.}~\bibnamefont
  {Yoshikawa}}, \bibinfo {author} {\bibfnamefont {S.}~\bibnamefont {Yokoyama}},
  \bibinfo {author} {\bibfnamefont {T.}~\bibnamefont {Kaji}}, \bibinfo {author}
  {\bibfnamefont {C.}~\bibnamefont {Sornphiphatphong}}, \bibinfo {author}
  {\bibfnamefont {Y.}~\bibnamefont {Shiozawa}}, \bibinfo {author}
  {\bibfnamefont {K.}~\bibnamefont {Makino}}, \ and\ \bibinfo {author}
  {\bibfnamefont {A.}~\bibnamefont {Furusawa}},\ }\href {\doibase
  10.1063/1.4962732} {\bibfield  {journal} {\bibinfo  {journal} {APL
  Photonics}\ }\textbf {\bibinfo {volume} {1}},\ \bibinfo {pages} {060801}
  (\bibinfo {year} {2016})}\BibitemShut {NoStop}%
\bibitem [{\citenamefont {Asavanant}\ \emph {et~al.}(2019)\citenamefont
  {Asavanant}, \citenamefont {Shiozawa}, \citenamefont {Yokoyama},
  \citenamefont {Charoensombutamon}, \citenamefont {Emura}, \citenamefont
  {Alexander}, \citenamefont {Takeda}, \citenamefont {Yoshikawa}, \citenamefont
  {Menicucci}, \citenamefont {Yonezawa},\ and\ \citenamefont
  {Furusawa}}]{Asavanant2019}%
  \BibitemOpen
  \bibfield  {author} {\bibinfo {author} {\bibfnamefont {W.}~\bibnamefont
  {Asavanant}}, \bibinfo {author} {\bibfnamefont {Y.}~\bibnamefont {Shiozawa}},
  \bibinfo {author} {\bibfnamefont {S.}~\bibnamefont {Yokoyama}}, \bibinfo
  {author} {\bibfnamefont {B.}~\bibnamefont {Charoensombutamon}}, \bibinfo
  {author} {\bibfnamefont {H.}~\bibnamefont {Emura}}, \bibinfo {author}
  {\bibfnamefont {R.~N.}\ \bibnamefont {Alexander}}, \bibinfo {author}
  {\bibfnamefont {S.}~\bibnamefont {Takeda}}, \bibinfo {author} {\bibfnamefont
  {J.}~\bibnamefont {Yoshikawa}}, \bibinfo {author} {\bibfnamefont {N.~C.}\
  \bibnamefont {Menicucci}}, \bibinfo {author} {\bibfnamefont {H.}~\bibnamefont
  {Yonezawa}}, \ and\ \bibinfo {author} {\bibfnamefont {A.}~\bibnamefont
  {Furusawa}},\ }\href {\doibase 10.1126/science.aay2645} {\bibfield  {journal}
  {\bibinfo  {journal} {Science}\ }\textbf {\bibinfo {volume} {366}},\ \bibinfo
  {pages} {373} (\bibinfo {year} {2019})}\BibitemShut {NoStop}%
\bibitem [{\citenamefont {Larsen}\ \emph {et~al.}(2019)\citenamefont {Larsen},
  \citenamefont {Guo}, \citenamefont {Breum}, \citenamefont
  {Neergaard-Nielsen},\ and\ \citenamefont {Andersen}}]{Larsen2019}%
  \BibitemOpen
  \bibfield  {author} {\bibinfo {author} {\bibfnamefont {M.~V.}\ \bibnamefont
  {Larsen}}, \bibinfo {author} {\bibfnamefont {X.}~\bibnamefont {Guo}},
  \bibinfo {author} {\bibfnamefont {C.~R.}\ \bibnamefont {Breum}}, \bibinfo
  {author} {\bibfnamefont {J.~S.}\ \bibnamefont {Neergaard-Nielsen}}, \ and\
  \bibinfo {author} {\bibfnamefont {U.~L.}\ \bibnamefont {Andersen}},\ }\href
  {\doibase 10.1126/science.aay4354} {\bibfield  {journal} {\bibinfo  {journal}
  {Science}\ }\textbf {\bibinfo {volume} {366}},\ \bibinfo {pages} {369}
  (\bibinfo {year} {2019})}\BibitemShut {NoStop}%
\bibitem [{\citenamefont {Asavanant}\ \emph {et~al.}()\citenamefont
  {Asavanant}, \citenamefont {Charoensombutamon}, \citenamefont {Yokoyama},
  \citenamefont {Ebihara}, \citenamefont {Nakamura}, \citenamefont {Alexander},
  \citenamefont {Endo}, \citenamefont {Yoshikawa}, \citenamefont {Menicucci},
  \citenamefont {Yonezawa},\ and\ \citenamefont {Furusawa}}]{Asavanant2020}%
  \BibitemOpen
  \bibfield  {author} {\bibinfo {author} {\bibfnamefont {W.}~\bibnamefont
  {Asavanant}}, \bibinfo {author} {\bibfnamefont {B.}~\bibnamefont
  {Charoensombutamon}}, \bibinfo {author} {\bibfnamefont {S.}~\bibnamefont
  {Yokoyama}}, \bibinfo {author} {\bibfnamefont {T.}~\bibnamefont {Ebihara}},
  \bibinfo {author} {\bibfnamefont {T.}~\bibnamefont {Nakamura}}, \bibinfo
  {author} {\bibfnamefont {R.~N.}\ \bibnamefont {Alexander}}, \bibinfo {author}
  {\bibfnamefont {M.}~\bibnamefont {Endo}}, \bibinfo {author} {\bibfnamefont
  {J.}~\bibnamefont {Yoshikawa}}, \bibinfo {author} {\bibfnamefont {N.~C.}\
  \bibnamefont {Menicucci}}, \bibinfo {author} {\bibfnamefont {H.}~\bibnamefont
  {Yonezawa}}, \ and\ \bibinfo {author} {\bibfnamefont {A.}~\bibnamefont
  {Furusawa}},\ }\href {http://arxiv.org/abs/2006.11537} {\ }\Eprint
  {http://arxiv.org/abs/2006.11537} {arXiv:2006.11537 [quant-ph]} \BibitemShut
  {NoStop}%
\bibitem [{\citenamefont {Alexander}\ \emph {et~al.}(2018)\citenamefont
  {Alexander}, \citenamefont {Yokoyama}, \citenamefont {Furusawa},\ and\
  \citenamefont {Menicucci}}]{Alexander2018}%
  \BibitemOpen
  \bibfield  {author} {\bibinfo {author} {\bibfnamefont {R.~N.}\ \bibnamefont
  {Alexander}}, \bibinfo {author} {\bibfnamefont {S.}~\bibnamefont {Yokoyama}},
  \bibinfo {author} {\bibfnamefont {A.}~\bibnamefont {Furusawa}}, \ and\
  \bibinfo {author} {\bibfnamefont {N.~C.}\ \bibnamefont {Menicucci}},\ }\href
  {\doibase 10.1103/PhysRevA.97.032302} {\bibfield  {journal} {\bibinfo
  {journal} {Physical Review A}\ }\textbf {\bibinfo {volume} {97}},\ \bibinfo
  {pages} {032302} (\bibinfo {year} {2018})}\BibitemShut {NoStop}%
\bibitem [{\citenamefont {Miyata}\ \emph {et~al.}(2016)\citenamefont {Miyata},
  \citenamefont {Ogawa}, \citenamefont {Marek}, \citenamefont {Filip},
  \citenamefont {Yonezawa}, \citenamefont {Yoshikawa},\ and\ \citenamefont
  {Furusawa}}]{Miyata2016}%
  \BibitemOpen
  \bibfield  {author} {\bibinfo {author} {\bibfnamefont {K.}~\bibnamefont
  {Miyata}}, \bibinfo {author} {\bibfnamefont {H.}~\bibnamefont {Ogawa}},
  \bibinfo {author} {\bibfnamefont {P.}~\bibnamefont {Marek}}, \bibinfo
  {author} {\bibfnamefont {R.}~\bibnamefont {Filip}}, \bibinfo {author}
  {\bibfnamefont {H.}~\bibnamefont {Yonezawa}}, \bibinfo {author}
  {\bibfnamefont {J.}~\bibnamefont {Yoshikawa}}, \ and\ \bibinfo {author}
  {\bibfnamefont {A.}~\bibnamefont {Furusawa}},\ }\href {\doibase
  10.1103/PhysRevA.93.022301} {\bibfield  {journal} {\bibinfo  {journal}
  {Physical Review A}\ }\textbf {\bibinfo {volume} {93}},\ \bibinfo {pages}
  {022301} (\bibinfo {year} {2016})}\BibitemShut {NoStop}%
\bibitem [{\citenamefont {Gottesman}\ \emph {et~al.}(2001)\citenamefont
  {Gottesman}, \citenamefont {Kitaev},\ and\ \citenamefont
  {Preskill}}]{Gottesman2001}%
  \BibitemOpen
  \bibfield  {author} {\bibinfo {author} {\bibfnamefont {D.}~\bibnamefont
  {Gottesman}}, \bibinfo {author} {\bibfnamefont {A.}~\bibnamefont {Kitaev}}, \
  and\ \bibinfo {author} {\bibfnamefont {J.}~\bibnamefont {Preskill}},\ }\href
  {\doibase 10.1103/PhysRevA.64.012310} {\bibfield  {journal} {\bibinfo
  {journal} {Physical Review A}\ }\textbf {\bibinfo {volume} {64}},\ \bibinfo
  {pages} {123101} (\bibinfo {year} {2001})}\BibitemShut {NoStop}%
\bibitem [{\citenamefont {Lvovsky}\ \emph {et~al.}(2009)\citenamefont
  {Lvovsky}, \citenamefont {Sanders},\ and\ \citenamefont
  {Tittel}}]{Lvovsky2009}%
  \BibitemOpen
  \bibfield  {author} {\bibinfo {author} {\bibfnamefont {A.~I.}\ \bibnamefont
  {Lvovsky}}, \bibinfo {author} {\bibfnamefont {B.~C.}\ \bibnamefont
  {Sanders}}, \ and\ \bibinfo {author} {\bibfnamefont {W.}~\bibnamefont
  {Tittel}},\ }\href {\doibase 10.1038/nphoton.2009.231} {\bibfield  {journal}
  {\bibinfo  {journal} {Nature Photonics}\ }\textbf {\bibinfo {volume} {3}},\
  \bibinfo {pages} {706} (\bibinfo {year} {2009})}\BibitemShut {NoStop}%
\bibitem [{\citenamefont {Lai}\ \emph {et~al.}(2018)\citenamefont {Lai},
  \citenamefont {Lin}, \citenamefont {Twamley},\ and\ \citenamefont
  {Goan}}]{Lai2018}%
  \BibitemOpen
  \bibfield  {author} {\bibinfo {author} {\bibfnamefont {Y.~Y.}\ \bibnamefont
  {Lai}}, \bibinfo {author} {\bibfnamefont {G.~D.}\ \bibnamefont {Lin}},
  \bibinfo {author} {\bibfnamefont {J.}~\bibnamefont {Twamley}}, \ and\
  \bibinfo {author} {\bibfnamefont {H.~S.}\ \bibnamefont {Goan}},\ }\href
  {\doibase 10.1103/PhysRevA.97.052303} {\bibfield  {journal} {\bibinfo
  {journal} {Physical Review A}\ }\textbf {\bibinfo {volume} {97}},\ \bibinfo
  {pages} {052303} (\bibinfo {year} {2018})}\BibitemShut {NoStop}%
\bibitem [{\citenamefont {Clausen}\ \emph {et~al.}(2011)\citenamefont
  {Clausen}, \citenamefont {Usmani}, \citenamefont {Bussi{\`{e}}res},
  \citenamefont {Sangouard}, \citenamefont {Afzelius}, \citenamefont
  {de~Riedmatten},\ and\ \citenamefont {Gisin}}]{Clausen2011}%
  \BibitemOpen
  \bibfield  {author} {\bibinfo {author} {\bibfnamefont {C.}~\bibnamefont
  {Clausen}}, \bibinfo {author} {\bibfnamefont {I.}~\bibnamefont {Usmani}},
  \bibinfo {author} {\bibfnamefont {F.}~\bibnamefont {Bussi{\`{e}}res}},
  \bibinfo {author} {\bibfnamefont {N.}~\bibnamefont {Sangouard}}, \bibinfo
  {author} {\bibfnamefont {M.}~\bibnamefont {Afzelius}}, \bibinfo {author}
  {\bibfnamefont {H.}~\bibnamefont {de~Riedmatten}}, \ and\ \bibinfo {author}
  {\bibfnamefont {N.}~\bibnamefont {Gisin}},\ }\href {\doibase
  10.1038/nature09662} {\bibfield  {journal} {\bibinfo  {journal} {Nature}\
  }\textbf {\bibinfo {volume} {469}},\ \bibinfo {pages} {508} (\bibinfo {year}
  {2011})}\BibitemShut {NoStop}%
\bibitem [{\citenamefont {Bimbard}\ \emph {et~al.}(2014)\citenamefont
  {Bimbard}, \citenamefont {Boddeda}, \citenamefont {Vitrant}, \citenamefont
  {Grankin}, \citenamefont {Parigi}, \citenamefont {Stanojevic}, \citenamefont
  {Ourjoumtsev},\ and\ \citenamefont {Grangier}}]{Bimbard2014}%
  \BibitemOpen
  \bibfield  {author} {\bibinfo {author} {\bibfnamefont {E.}~\bibnamefont
  {Bimbard}}, \bibinfo {author} {\bibfnamefont {R.}~\bibnamefont {Boddeda}},
  \bibinfo {author} {\bibfnamefont {N.}~\bibnamefont {Vitrant}}, \bibinfo
  {author} {\bibfnamefont {A.}~\bibnamefont {Grankin}}, \bibinfo {author}
  {\bibfnamefont {V.}~\bibnamefont {Parigi}}, \bibinfo {author} {\bibfnamefont
  {J.}~\bibnamefont {Stanojevic}}, \bibinfo {author} {\bibfnamefont
  {A.}~\bibnamefont {Ourjoumtsev}}, \ and\ \bibinfo {author} {\bibfnamefont
  {P.}~\bibnamefont {Grangier}},\ }\href {\doibase
  10.1103/PhysRevLett.112.033601} {\bibfield  {journal} {\bibinfo  {journal}
  {Physical Review Letters}\ }\textbf {\bibinfo {volume} {112}},\ \bibinfo
  {pages} {033601} (\bibinfo {year} {2014})}\BibitemShut {NoStop}%
\bibitem [{\citenamefont {Bouillard}\ \emph {et~al.}(2019)\citenamefont
  {Bouillard}, \citenamefont {Boucher}, \citenamefont {{Ferrer Ortas}},
  \citenamefont {Pointard},\ and\ \citenamefont
  {Tualle-Brouri}}]{Bouillard2019}%
  \BibitemOpen
  \bibfield  {author} {\bibinfo {author} {\bibfnamefont {M.}~\bibnamefont
  {Bouillard}}, \bibinfo {author} {\bibfnamefont {G.}~\bibnamefont {Boucher}},
  \bibinfo {author} {\bibfnamefont {J.}~\bibnamefont {{Ferrer Ortas}}},
  \bibinfo {author} {\bibfnamefont {B.}~\bibnamefont {Pointard}}, \ and\
  \bibinfo {author} {\bibfnamefont {R.}~\bibnamefont {Tualle-Brouri}},\ }\href
  {\doibase 10.1103/PhysRevLett.122.210501} {\bibfield  {journal} {\bibinfo
  {journal} {Physical Review Letters}\ }\textbf {\bibinfo {volume} {122}},\
  \bibinfo {pages} {210501} (\bibinfo {year} {2019})}\BibitemShut {NoStop}%
\bibitem [{\citenamefont {Yoshikawa}\ \emph {et~al.}(2013)\citenamefont
  {Yoshikawa}, \citenamefont {Makino}, \citenamefont {Kurata}, \citenamefont
  {van Loock},\ and\ \citenamefont {Furusawa}}]{Yoshikawa2014}%
  \BibitemOpen
  \bibfield  {author} {\bibinfo {author} {\bibfnamefont {J.}~\bibnamefont
  {Yoshikawa}}, \bibinfo {author} {\bibfnamefont {K.}~\bibnamefont {Makino}},
  \bibinfo {author} {\bibfnamefont {S.}~\bibnamefont {Kurata}}, \bibinfo
  {author} {\bibfnamefont {P.}~\bibnamefont {van Loock}}, \ and\ \bibinfo
  {author} {\bibfnamefont {A.}~\bibnamefont {Furusawa}},\ }\href {\doibase
  10.1103/PhysRevX.3.041028} {\bibfield  {journal} {\bibinfo  {journal}
  {Physical Review X}\ }\textbf {\bibinfo {volume} {3}},\ \bibinfo {pages}
  {041028} (\bibinfo {year} {2013})}\BibitemShut {NoStop}%
\bibitem [{\citenamefont {Hashimoto}\ \emph {et~al.}(2019)\citenamefont
  {Hashimoto}, \citenamefont {Toyama}, \citenamefont {Yoshikawa}, \citenamefont
  {Makino}, \citenamefont {Okamoto}, \citenamefont {Sakakibara}, \citenamefont
  {Takeda}, \citenamefont {{van Loock}},\ and\ \citenamefont
  {Furusawa}}]{Hashimoto2019}%
  \BibitemOpen
  \bibfield  {author} {\bibinfo {author} {\bibfnamefont {Y.}~\bibnamefont
  {Hashimoto}}, \bibinfo {author} {\bibfnamefont {T.}~\bibnamefont {Toyama}},
  \bibinfo {author} {\bibfnamefont {J.}~\bibnamefont {Yoshikawa}}, \bibinfo
  {author} {\bibfnamefont {K.}~\bibnamefont {Makino}}, \bibinfo {author}
  {\bibfnamefont {F.}~\bibnamefont {Okamoto}}, \bibinfo {author} {\bibfnamefont
  {R.}~\bibnamefont {Sakakibara}}, \bibinfo {author} {\bibfnamefont
  {S.}~\bibnamefont {Takeda}}, \bibinfo {author} {\bibfnamefont
  {P.}~\bibnamefont {{van Loock}}}, \ and\ \bibinfo {author} {\bibfnamefont
  {A.}~\bibnamefont {Furusawa}},\ }\href {\doibase
  10.1103/PhysRevLett.123.113603} {\bibfield  {journal} {\bibinfo  {journal}
  {Physical Review Letters}\ }\textbf {\bibinfo {volume} {123}},\ \bibinfo
  {pages} {113603} (\bibinfo {year} {2019})}\BibitemShut {NoStop}%
\bibitem [{\citenamefont {Makino}\ \emph {et~al.}(2016)\citenamefont {Makino},
  \citenamefont {Hashimoto}, \citenamefont {Yoshikawa}, \citenamefont {Ohdan},
  \citenamefont {Toyama}, \citenamefont {{van Loock}},\ and\ \citenamefont
  {Furusawa}}]{Makino2016}%
  \BibitemOpen
  \bibfield  {author} {\bibinfo {author} {\bibfnamefont {K.}~\bibnamefont
  {Makino}}, \bibinfo {author} {\bibfnamefont {Y.}~\bibnamefont {Hashimoto}},
  \bibinfo {author} {\bibfnamefont {J.}~\bibnamefont {Yoshikawa}}, \bibinfo
  {author} {\bibfnamefont {H.}~\bibnamefont {Ohdan}}, \bibinfo {author}
  {\bibfnamefont {T.}~\bibnamefont {Toyama}}, \bibinfo {author} {\bibfnamefont
  {P.}~\bibnamefont {{van Loock}}}, \ and\ \bibinfo {author} {\bibfnamefont
  {A.}~\bibnamefont {Furusawa}},\ }\href {\doibase 10.1126/sciadv.1501772}
  {\bibfield  {journal} {\bibinfo  {journal} {Science Advances}\ }\textbf
  {\bibinfo {volume} {2}},\ \bibinfo {pages} {e1501772} (\bibinfo {year}
  {2016})}\BibitemShut {NoStop}%
\bibitem [{Sup()}]{Supplemental}%
  \BibitemOpen
  \href@noop {} {\bibinfo  {journal} {See Supplemental Material at [URL will be
  inserted by publisher] for the methods, the preliminary experiment, the
  experimental results with various parameters and the precise calculations for
  the losses, which includes Refs. [15-16,25,26]}\ }\BibitemShut {NoStop}%
\bibitem [{\citenamefont {Yoshikawa}\ \emph {et~al.}(2018)\citenamefont
  {Yoshikawa}, \citenamefont {Bergmann}, \citenamefont {{van Loock}},
  \citenamefont {Fuwa}, \citenamefont {Okada}, \citenamefont {Takase},
  \citenamefont {Toyama}, \citenamefont {Makino}, \citenamefont {Takeda},\ and\
  \citenamefont {Furusawa}}]{Yoshikawa2018}%
  \BibitemOpen
\bibfield  {journal} {  }\bibfield  {author} {\bibinfo {author} {\bibfnamefont
  {J.}~\bibnamefont {Yoshikawa}}, \bibinfo {author} {\bibfnamefont
  {M.}~\bibnamefont {Bergmann}}, \bibinfo {author} {\bibfnamefont
  {P.}~\bibnamefont {{van Loock}}}, \bibinfo {author} {\bibfnamefont
  {M.}~\bibnamefont {Fuwa}}, \bibinfo {author} {\bibfnamefont {M.}~\bibnamefont
  {Okada}}, \bibinfo {author} {\bibfnamefont {K.}~\bibnamefont {Takase}},
  \bibinfo {author} {\bibfnamefont {T.}~\bibnamefont {Toyama}}, \bibinfo
  {author} {\bibfnamefont {K.}~\bibnamefont {Makino}}, \bibinfo {author}
  {\bibfnamefont {S.}~\bibnamefont {Takeda}}, \ and\ \bibinfo {author}
  {\bibfnamefont {A.}~\bibnamefont {Furusawa}},\ }\href {\doibase
  10.1103/PhysRevA.97.053814} {\bibfield  {journal} {\bibinfo  {journal}
  {Physical Review A}\ }\textbf {\bibinfo {volume} {97}},\ \bibinfo {pages}
  {053814} (\bibinfo {year} {2018})}\BibitemShut {NoStop}%
\bibitem [{\citenamefont {Schwinger}(1952)}]{JulianSeymourSchwinger}%
  \BibitemOpen
  \bibfield  {author} {\bibinfo {author} {\bibfnamefont {J.}~\bibnamefont
  {Schwinger}},\ }\href@noop {} {\emph {\bibinfo {title} {{ON ANGULAR
  MOMENTUM}}}},\ \bibinfo {type} {Tech. Rep.}\ (\bibinfo  {institution}
  {Harvard Univ., Cambridge, MA (United States); Nuclear Development
  Associates, Inc. (US)},\ \bibinfo {year} {1952})\BibitemShut {NoStop}%
\bibitem [{\citenamefont {Chuang}\ \emph {et~al.}(1997)\citenamefont {Chuang},
  \citenamefont {Leung},\ and\ \citenamefont {Yamamoto}}]{Chuang1997}%
  \BibitemOpen
  \bibfield  {author} {\bibinfo {author} {\bibfnamefont {I.~L.}\ \bibnamefont
  {Chuang}}, \bibinfo {author} {\bibfnamefont {D.~W.}\ \bibnamefont {Leung}}, \
  and\ \bibinfo {author} {\bibfnamefont {Y.}~\bibnamefont {Yamamoto}},\ }\href
  {\doibase 10.1103/PhysRevA.56.1114} {\bibfield  {journal} {\bibinfo
  {journal} {Physical Review A}\ }\textbf {\bibinfo {volume} {56}},\ \bibinfo
  {pages} {1114} (\bibinfo {year} {1997})}\BibitemShut {NoStop}%
\bibitem [{\citenamefont {Jin}\ \emph {et~al.}(2013)\citenamefont {Jin},
  \citenamefont {Peng}, \citenamefont {Deng}, \citenamefont {Barbieri},
  \citenamefont {Nunn},\ and\ \citenamefont {Walmsley}}]{Jin2013c}%
  \BibitemOpen
  \bibfield  {author} {\bibinfo {author} {\bibfnamefont {X.~M.}\ \bibnamefont
  {Jin}}, \bibinfo {author} {\bibfnamefont {C.~Z.}\ \bibnamefont {Peng}},
  \bibinfo {author} {\bibfnamefont {Y.}~\bibnamefont {Deng}}, \bibinfo {author}
  {\bibfnamefont {M.}~\bibnamefont {Barbieri}}, \bibinfo {author}
  {\bibfnamefont {J.}~\bibnamefont {Nunn}}, \ and\ \bibinfo {author}
  {\bibfnamefont {I.~A.}\ \bibnamefont {Walmsley}},\ }\href {\doibase
  10.1038/srep01779} {\bibfield  {journal} {\bibinfo  {journal} {Scientific
  Reports}\ }\textbf {\bibinfo {volume} {3}},\ \bibinfo {pages} {1779}
  (\bibinfo {year} {2013})}\BibitemShut {NoStop}%
\bibitem [{\citenamefont {M{\"{u}}ller}\ \emph {et~al.}(2017)\citenamefont
  {M{\"{u}}ller}, \citenamefont {Vural}, \citenamefont {Schneider},
  \citenamefont {Rastelli}, \citenamefont {Schmidt}, \citenamefont
  {H{\"{o}}fling},\ and\ \citenamefont {Michler}}]{Muller2017}%
  \BibitemOpen
  \bibfield  {author} {\bibinfo {author} {\bibfnamefont {M.}~\bibnamefont
  {M{\"{u}}ller}}, \bibinfo {author} {\bibfnamefont {H.}~\bibnamefont {Vural}},
  \bibinfo {author} {\bibfnamefont {C.}~\bibnamefont {Schneider}}, \bibinfo
  {author} {\bibfnamefont {A.}~\bibnamefont {Rastelli}}, \bibinfo {author}
  {\bibfnamefont {O.~G.}\ \bibnamefont {Schmidt}}, \bibinfo {author}
  {\bibfnamefont {S.}~\bibnamefont {H{\"{o}}fling}}, \ and\ \bibinfo {author}
  {\bibfnamefont {P.}~\bibnamefont {Michler}},\ }\href {\doibase
  10.1103/PhysRevLett.118.257402} {\bibfield  {journal} {\bibinfo  {journal}
  {Physical Review Letters}\ }\textbf {\bibinfo {volume} {118}},\ \bibinfo
  {pages} {257402} (\bibinfo {year} {2017})}\BibitemShut {NoStop}%
\bibitem [{\citenamefont {Bergmann}\ and\ \citenamefont {{van
  Loock}}(2016)}]{Bergmann2016}%
  \BibitemOpen
  \bibfield  {author} {\bibinfo {author} {\bibfnamefont {M.}~\bibnamefont
  {Bergmann}}\ and\ \bibinfo {author} {\bibfnamefont {P.}~\bibnamefont {{van
  Loock}}},\ }\href {\doibase 10.1103/PhysRevA.94.012311} {\bibfield  {journal}
  {\bibinfo  {journal} {Physical Review A}\ }\textbf {\bibinfo {volume} {94}},\
  \bibinfo {pages} {012311} (\bibinfo {year} {2016})}\BibitemShut {NoStop}%
\bibitem [{\citenamefont {Macrae}\ \emph {et~al.}(2012)\citenamefont {Macrae},
  \citenamefont {Brannan}, \citenamefont {Achal},\ and\ \citenamefont
  {Lvovsky}}]{Macrae2012}%
  \BibitemOpen
  \bibfield  {author} {\bibinfo {author} {\bibfnamefont {A.}~\bibnamefont
  {Macrae}}, \bibinfo {author} {\bibfnamefont {T.}~\bibnamefont {Brannan}},
  \bibinfo {author} {\bibfnamefont {R.}~\bibnamefont {Achal}}, \ and\ \bibinfo
  {author} {\bibfnamefont {A.~I.}\ \bibnamefont {Lvovsky}},\ }\href {\doibase
  10.1103/PhysRevLett.109.033601} {\bibfield  {journal} {\bibinfo  {journal}
  {Physical Review Letters}\ }\textbf {\bibinfo {volume} {109}},\ \bibinfo
  {pages} {033601} (\bibinfo {year} {2012})}\BibitemShut {NoStop}%
\bibitem [{\citenamefont {Morin}\ \emph {et~al.}(2013)\citenamefont {Morin},
  \citenamefont {Fabre},\ and\ \citenamefont {Laurat}}]{Morin2013}%
  \BibitemOpen
  \bibfield  {author} {\bibinfo {author} {\bibfnamefont {O.}~\bibnamefont
  {Morin}}, \bibinfo {author} {\bibfnamefont {C.}~\bibnamefont {Fabre}}, \ and\
  \bibinfo {author} {\bibfnamefont {J.}~\bibnamefont {Laurat}},\ }\href
  {\doibase 10.1103/PhysRevLett.111.213602} {\bibfield  {journal} {\bibinfo
  {journal} {Physical Review Letters}\ }\textbf {\bibinfo {volume} {111}},\
  \bibinfo {pages} {213602} (\bibinfo {year} {2013})}\BibitemShut {NoStop}%
\bibitem [{\citenamefont {Plenio}(2005)}]{Plenio2005}%
  \BibitemOpen
  \bibfield  {author} {\bibinfo {author} {\bibfnamefont {M.~B.}\ \bibnamefont
  {Plenio}},\ }\href {\doibase 10.1103/PhysRevLett.95.090503} {\bibfield
  {journal} {\bibinfo  {journal} {Physical Review Letters}\ }\textbf {\bibinfo
  {volume} {95}},\ \bibinfo {pages} {090503} (\bibinfo {year}
  {2005})}\BibitemShut {NoStop}%
\bibitem [{\citenamefont {Vidal}\ and\ \citenamefont
  {Werner}(2002)}]{Vidal2002}%
  \BibitemOpen
  \bibfield  {author} {\bibinfo {author} {\bibfnamefont {G.}~\bibnamefont
  {Vidal}}\ and\ \bibinfo {author} {\bibfnamefont {R.~F.}\ \bibnamefont
  {Werner}},\ }\href {\doibase 10.1103/PhysRevA.65.032314} {\bibfield
  {journal} {\bibinfo  {journal} {Physical Review A}\ }\textbf {\bibinfo
  {volume} {65}},\ \bibinfo {pages} {032314} (\bibinfo {year}
  {2002})}\BibitemShut {NoStop}%
\bibitem [{\citenamefont {Eisert}(2001)}]{Plenio1998}%
  \BibitemOpen
  \bibfield  {author} {\bibinfo {author} {\bibfnamefont {J.}~\bibnamefont
  {Eisert}},\ }\emph {\bibinfo {title} {{Entanglement in Quantum Information
  Theory}}},\ \href {\doibase 10.1080/001075198181766} {Ph.D. thesis},\
  \bibinfo  {school} {University of Potsdam, Germany} (\bibinfo {year}
  {2001})\BibitemShut {NoStop}%
\end{thebibliography}%

%%%%%%%%%% Merge with supplemental materials %%%%%%%%%%
\pagebreak
\widetext
\begin{center}
\textbf{\large Supplementary material: Phase Locking between Two All-Optical Quantum Memories}
\end{center}
%%%%%%%%%% Merge with supplemental materials %%%%%%%%%%
%%%%%%%%%% Prefix a "S" to all equations, figures, tables and reset the counter %%%%%%%%%%
\setcounter{equation}{0}
\setcounter{figure}{0}
\setcounter{table}{0}
\setcounter{page}{1}
\makeatletter
\renewcommand{\theequation}{S\arabic{equation}}
\renewcommand{\thefigure}{S\arabic{figure}}
\renewcommand{\bibnumfmt}[1]{[S#1]}
\renewcommand{\citenumfont}[1]{S#1}

\maketitle

\section{Methods}
Our experimental apparatus is shown in Fig.~3 of the main text.
The light source is a continuous-wave Ti:Sapphire laser operating at the wavelength of 860 nm (MBR-110, Coherent).
The second harmonics with the power of about 2.0 mW and 1.7 mW are used as pump beams to create photon pairs inside memory cavities 1 and 2. 
The memory cavity, containing a periodically-poled KTiOPO$_4$ (PPKTP) crystal (Raicol) as a nonlinear optical medium, has a round-trip length of about 1.4 m, while the shutter cavity, containing an RbTiOPO$_4$(RTP) EOM (Leysop), has that of about 0.7 m. 
The PPKTP crystal is type-0 quasi-phase-matched and has a length of 10 mm. 
The reflectivity of the coupling mirror between the memory and the shutter cavities is about 98\%, and that of the outcoupling mirror of the shutter is about 72\%. 
The EOM in the shutter is driven by a high-voltage switch (Bergmann Messger\"ate Entwicklung). 
The latency of the high-voltage switch is less than 50 ns, and the applied voltage is estimated as about 900 V.

Three filter cavities are applied to the idler field. 
The first one is bow-tie shaped and has a round-trip length of about 250 mm (referred to as separating cavity in the main text).  
The signal field is reflected by this separating cavity and directed to a homodyne detector for the state characterization, while the idler field is transmitted through the cavity. 
Two additional cavities are Fabry-Perot cavities for the idler fields. 
A single tooth of the frequency comb of the memory cavity is selected by these cavities and directed to a photon detector. 
The idler beams to determine the entangled states are combined with each other at a beam splitter between the first and the second filter cavities.
The photon detector contains a silicon avalanche photodiode (APD, SPCM-AQRH-16-FC, Excelitas Technologies) and is coupled with an optical fiber. 
The heralding event rate depends on the experimental condition, and a typical rate for balanced conditions $(\ket{0,1} + \ket{1,0})/\sqrt{2}$ is about 400 counts per second (without compensation of the duty cycle).

Several AOSs and AOFSs are used in the experiment, as depicted in Fig.~3. 
AOS-1 and AOS-2 are for chopping the probe beams. 
Driving signals for these AOSs are on when the optical systems are feedback controlled, and diffracted beams are used as probe beams, while they are off when the entangled states are created. 
AOFS-1 and AOFS-2 are for shifting the laser frequency by the free spectral range of the memory cavity.  
AOS-5 is inserted for protecting the APD from the idler probe beam, by coupling a diffracted beam to the APD and switching the driving signal. 
The detuning of the signal mode from the LO frequency, explained in the main text, is controlled via the difference of the driving frequencies between AOS-1 and AOS-2.
The signal probe beam is detuned by 200 kHz, and by locking the memory cavity to almost the bottom of the fringe in the transmitted signal probe beam, the total detuning of the signal mode by about 300 kHz from the LO is stably realized. 
Frequencies and phases of the driving signals for the AOSs and AOFSs are precisely controlled by direct digital synthesizers (AD9959, Analog Devices Inc.).

Even though several detunings are employed in the experiment, the optical frequencies of each signal mode are almost the same.
Experimentally, this means that the frequencies of the lock beams for the OPOs are almost the same.
Thanks to this, the phase rotation of the state shown in the previous experiment [16] is distinguished when each mode is emitted simultaneously, while the rotation remains when there are differences for the emission timing of each mode.

Figure~3 is somewhat simplified from the actual apparatus for simplicity. 
From Fig.~3, some beams for controlling the optical systems are omitted, which we call ``lock beams'' in the following. 
The lock beams are periodically chopped in the same way as the probe beams. 
For example, for each cavity, a lock beam is employed for the purpose of obtaining an error signal for locking the cavity length. 
As for the memory cavity, first a lock beam is used for roughly locking the cavity length, and after the rough locking, next the transmitted signal probe beam is used for precisely locking the detuning frequency. 
The lock beam is injected into the memory cavity with vertical polarization, while the pump, signal and idler fields are in horizontal polarization.

%%%%%%%%%%%%%%%%
\section{Storage of single-photon state}
As a preliminary experiment, we tested storage and release of single-photon states.
In this case, the beam from one memory is blocked, while the other experimental elements are the same as for the experiment with entangled states.
In addition, we replace the 50:50 beamsplitter for combining the idler beams to a HR or anti-reflection coated mirror.
This part is almost the same as in our former experiments [15-17], the difference, however, is that now the probe beams are depending on which a phase-sensitive full quantum tomography is conducted and the idler beams are combined before the APD, whereas in the former experiments the assumed quantum states are either phase insensitive or only mode.

This preliminary experiment is performed in order to obtain wave packet (WP) envelope functions $\Psi(t)$ of the released quantum states for various storage times by the principal component analysis [15,25,26]. 
Quadrature values of the target quantum states are obtained from continuous homodyne signals, denoted by $\hat{x}(t)$, via weighted integration $\hat{x}=\int \text{d}t\Psi(t)\hat{x}(t)$. 
The functions $\Psi(t)$ obtained in this preliminary experiment are also used for the characterization of entangled states.

\begin{figure}[tbp]
\centering
\includegraphics[scale = 0.8,pagebox=artbox]{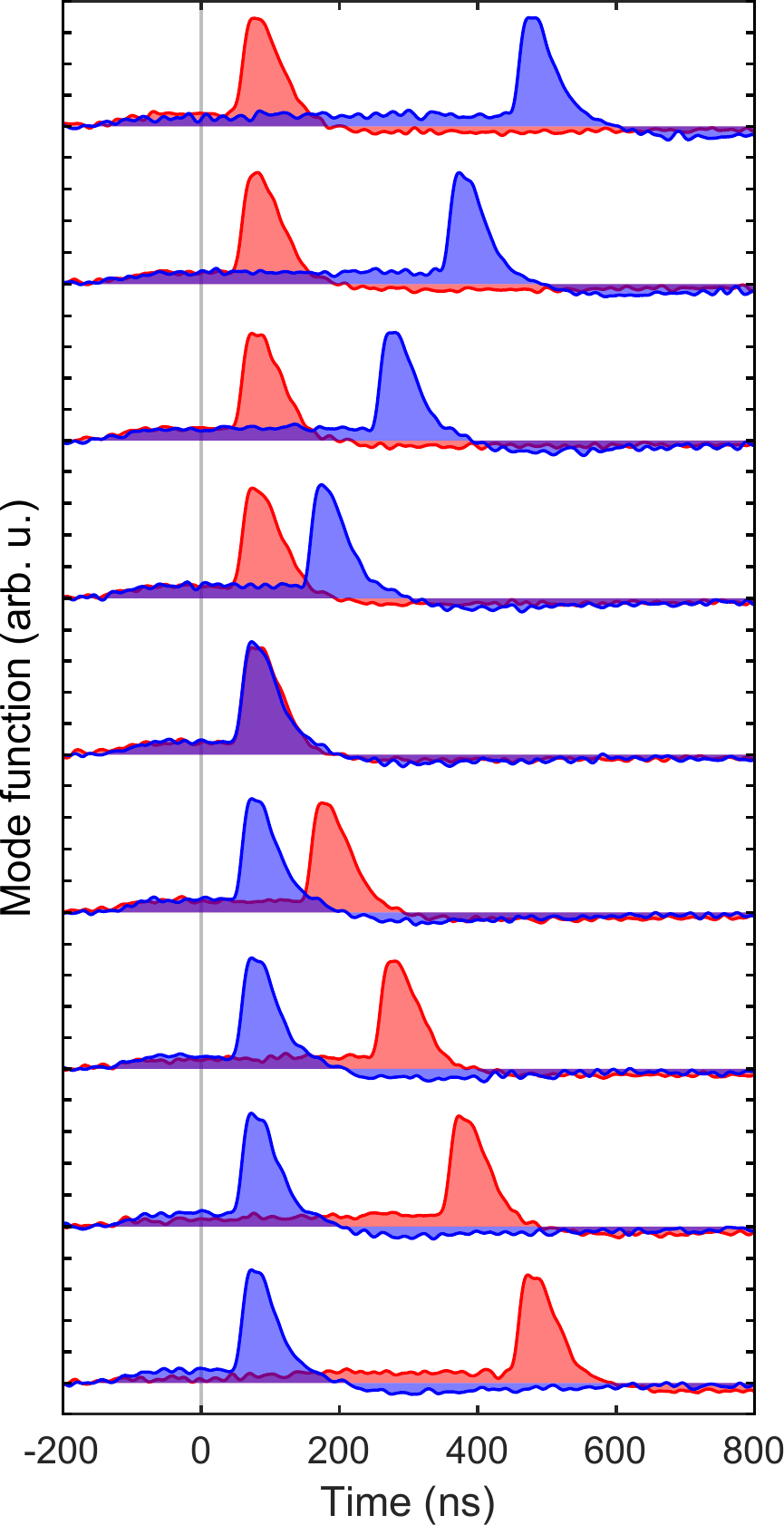}
\caption{Storage and release of single-photon states.
Red: WP envelope functions $\Psi(t)$ of the released states of CCS-1.
Blue: WP envelope functions $\Psi(t)$ of the released states of CCS-2.
The time origin corresponds to timings of heralding signals.
}
\label{fig:preliminary}
\end{figure}

\begin{figure}[tbp]
\centering
\includegraphics[scale = 0.8]{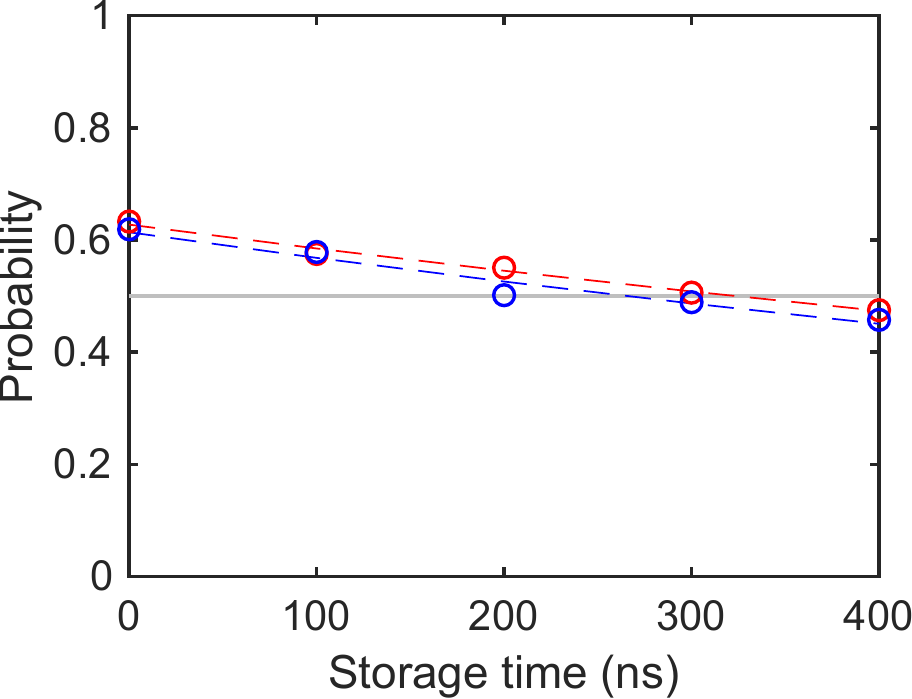}
\caption{Probabilities of single-photon states as a function of the storage time.
Red: single-photon fraction of CCS1.
Blue: single-photon fraction of CCS-2.
Circles show experimental values, whereas traces are fitted curves assuming exponential decay.
}
\label{fig:singlephoton}
\end{figure}

The results of the preliminary experiment are shown in Fig.~\ref{fig:preliminary}.
Figure~\ref{fig:preliminary} shows estimated WP envelope functions $\Psi(t)$ for storage time differences  between the two modes of 0 ns, 100 ns, 200 ns, 300 ns, and 400 ns. 
The horizontal axis is the relative time where the timing of the heralding signal corresponds to 0 ns. 
The shape of the WP is independent of the storage time, except for small leakage before release, and appropriately shifted in accordance with the storage time. 
Figure~\ref{fig:singlephoton} shows single-photon fractions for various storage times for each CCS. 

%%%%%%%%%%%%%%%%
\section{Simultaneous quadrature distributions}
\begin{figure*}
\includegraphics[width = 0.9\textwidth]{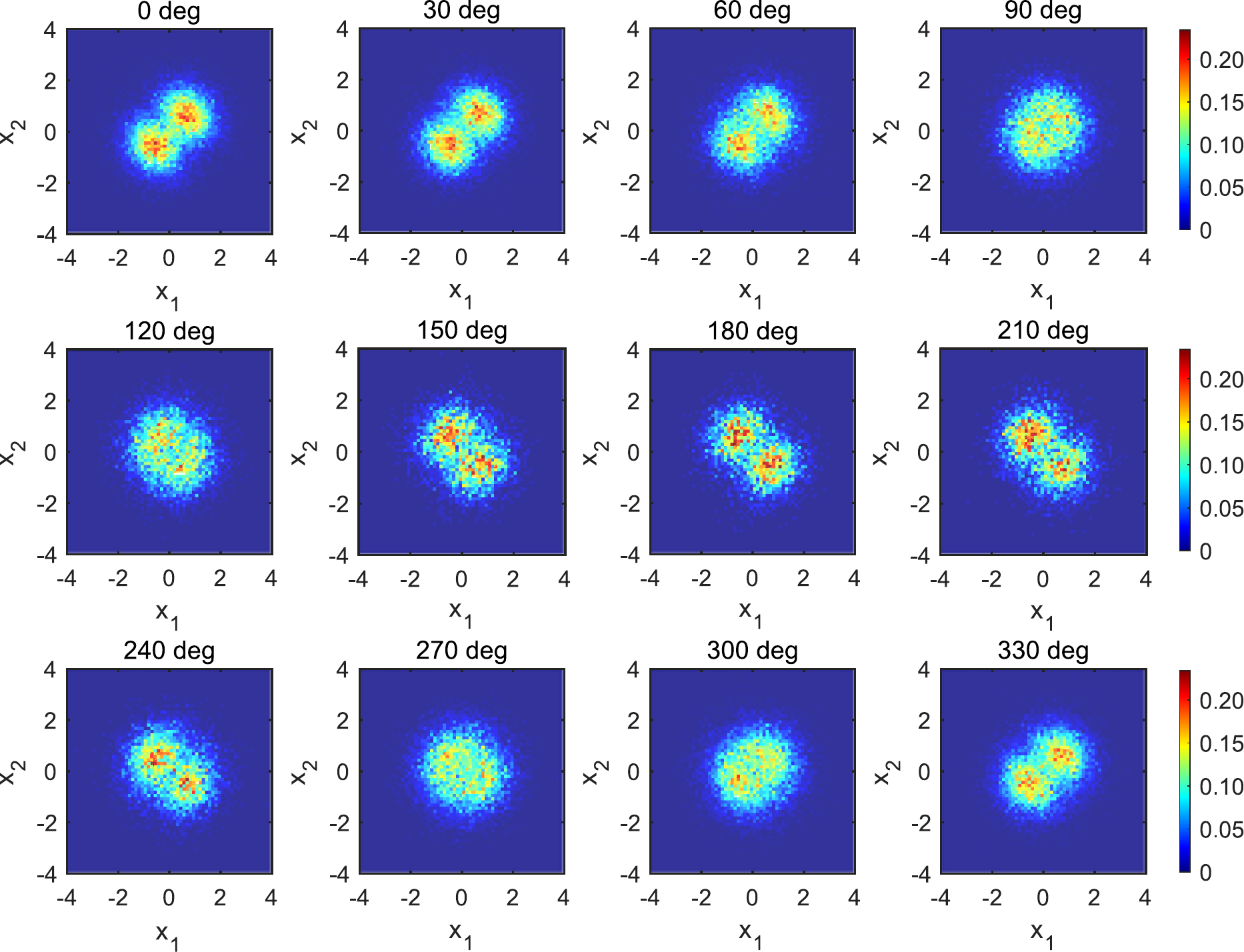}
\caption{Experimental simultaneous quadrature distributions of $(\ket{0,1} + \ket{1,0})/\sqrt{2}$ of 0 ns storage time for each relative homodyne phase.
First row: from left to right, the relative phase varies from 0$^\circ$ to 90$^\circ$.
Second row: from left to right, the relative phase varies from 120$^\circ$ to 210$^\circ$.
Third row: from left to right, the relative phase varies from 240$^\circ$ to 330$^\circ$.
}
\label{fig:storage_QR}
\end{figure*}

Here, we present experimental results of simultaneous quadrature distributions for various phases and density matrices.
Figure~\ref{fig:storage_QR} shows the experimentally obtained simultaneous quadrature distributions of two-mode homodyne measurements of $(\ket{0,1} + \ket{1,0})/\sqrt{2}$, where the storage time is set to 0 ns for each mode, so that the emission for each mode is done simultaneously.
We perform the balanced optical homodyne measurements at various local oscillators' (LOs') phases $\varphi_1, \varphi_2$ by homodyne detectors 1, 2, respectively.
The phases of LOs $\varphi_1, \varphi_2$ appear in the measurements like $\bra{x_1,x_2}\text{e}^{\text{i}(\varphi_1\hat{a}^\dagger_1\hat{a}_1+\varphi_2\hat{a}^\dagger_2\hat{a}_2)}$.
Then the probability distribution at the LOs' phases $\varphi_1, \varphi_2$ for the state $\ket{\psi} = \alpha\ket{0,1}+\beta \text{e}^{\text{i}\theta}\ket{1,0}$ becomes
\begin{align}
&|\braket{x_1,x_2|\text{e}^{\text{i}(\varphi_1\hat{a}^\dagger_1\hat{a}_1+\varphi_2\hat{a}^\dagger_2\hat{a}_2)}|\psi}|^2\nonumber\\
&=|\text{e}^{\text{i}\varphi_2}\bra{x_1, x_2}(\alpha\ket{0,1}+\beta \text{e}^{\text{i}(\varphi_1-\varphi_2+\theta)}\ket{1,0})|^2.
\end{align}
Therefore the relative phase of LOs $\Delta\varphi=\varphi_1-\varphi_2$ affects the experimentally obtained probability distributions like the phase rotation of the state.
When the phase becomes 0$^\circ$, the distribution exhibits two-peak shapes, and for a phase of 90$^\circ$, the distribution becomes circle shaped. Finally the distribution becomes two-peak shaped.% leaned other side than in 0$^\circ$ when the phase becomes 180$^\circ$.

\section{Phase rotation}
As mentioned in the main text,  the signal mode and the local oscillator are detuned by $\Delta\omega\sim2\pi\times300$~kHz by slightly tuning the driving frequencies for AOS-1 and AOS-2.
 This is utilized in order to avoid the back-scattered light of the local oscillator beams at the homodyne detectors stored in the memory cavities.
 This detuning $\Delta\omega$ causes the phase rotation of the quantum state.
 The phase rotation can be prevented by simultaneous emission of two modes, since the generated state is a photon-number fixed state (in our case, the total photon number is fixed to one).
 The rotation is described by $\text{e}^{\text{i}\Delta\omega t\hat{a}^\dagger\hat{a}}$, and then the state $\ket{\psi}$ is rotated as
 \begin{align}
 &\text{e}^{\text{i}\Delta\omega t_1\hat{a}^\dagger_1\hat{a}_1}\text{e}^{\text{i}\Delta\omega t_2\hat{a}^\dagger_2\hat{a}_2}\ket{\psi}\nonumber\\
 &= \text{e}^{\text{i}\Delta\omega t_2}(\alpha\ket{0,1}+\beta \text{e}^{\text{i}\theta}\text{e}^{-\text{i}\Delta\omega\Delta\tau}\ket{1,0}).
 \end{align}
 Since times other than the delay times are only involved in the global phase, only the time delay can affect the phase rotation caused by the detuning as it appeared in the previous work of [16].

%%%%%%%%%%%%%%%%
\section{Storing and releasing various two-mode entangled states}

\begin{figure*}
\includegraphics[width = 0.9\textwidth]{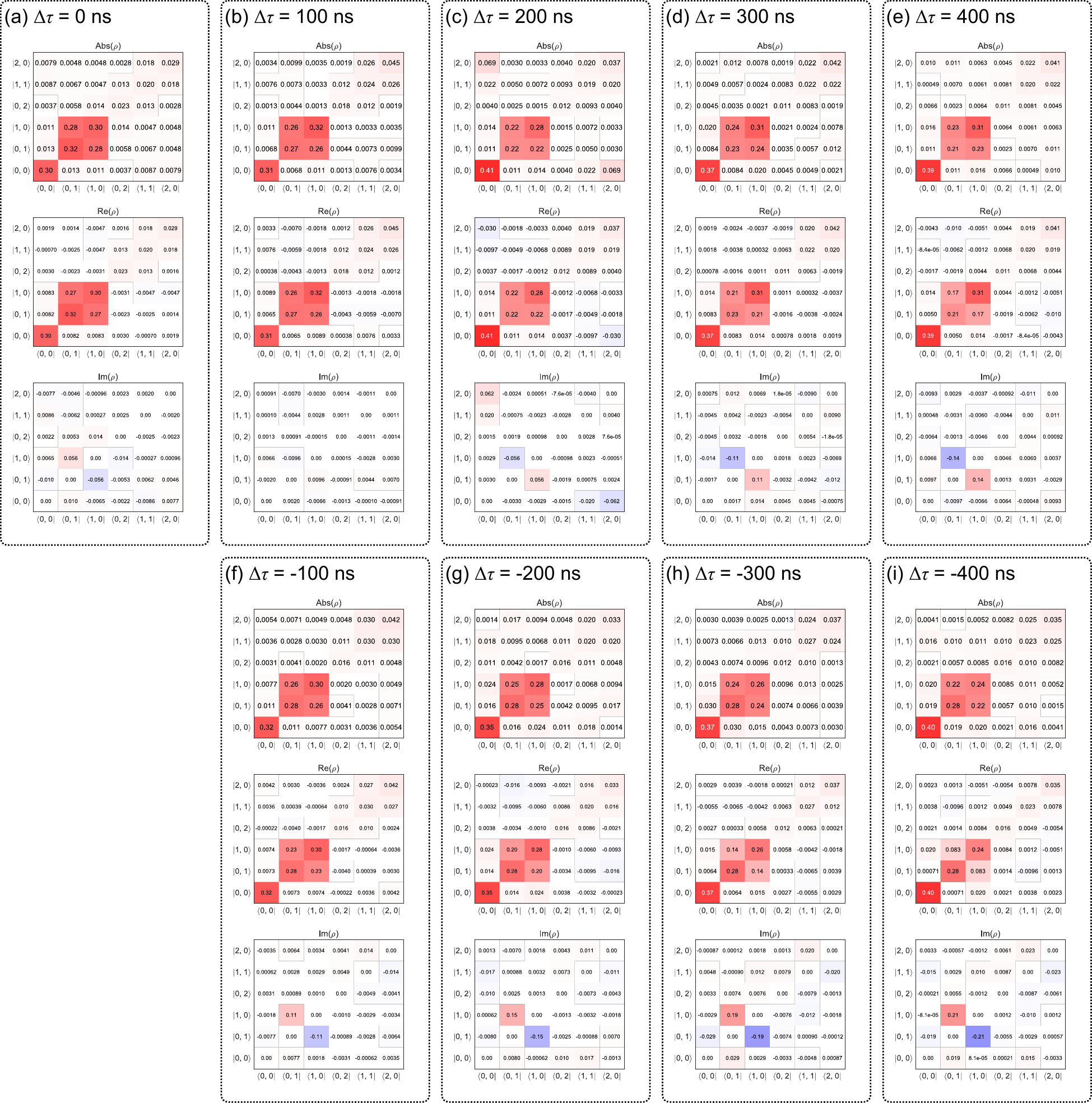}
\caption{Experimental results of storing $(\ket{0,1} + \ket{1,0})/\sqrt{2}$ with different emission timing for each mode.
(a-i)Upper row: absolute value of density matrices.
Middle row: real part of density matrices.
Lower row: imaginary part of density matrices.
}
\label{fig:storage_DMDelay}
\end{figure*}
\begin{figure}
\includegraphics[scale = 0.8]{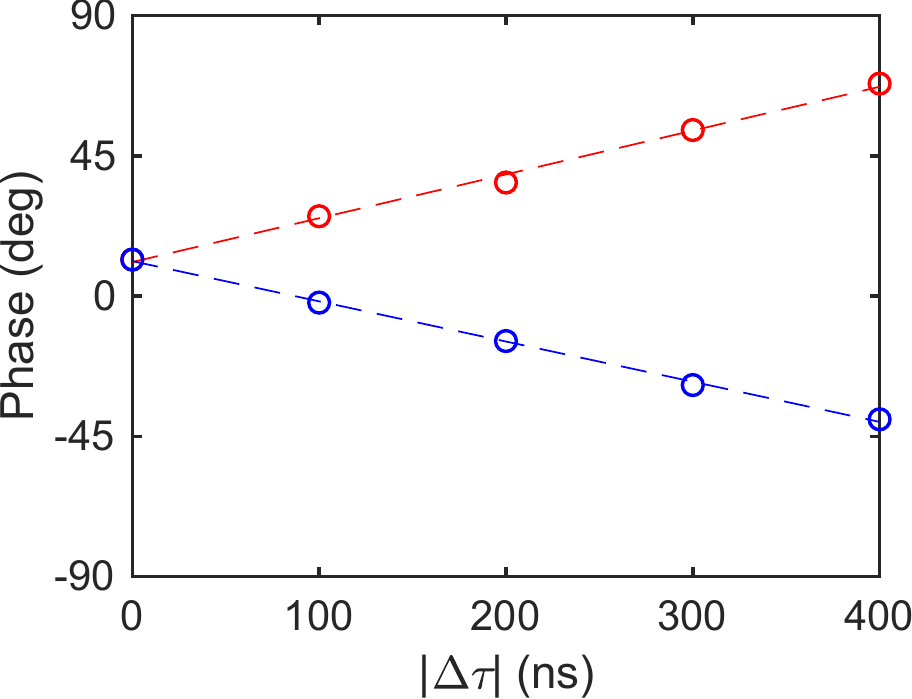}
\caption{Phases of experimentally obtained density matrices of $(\ket{0,1}+\ket{1,0})/\sqrt{2}$.
Red: Delayed release of mode-1.
Blue: Delayed release of mode-2.
%The rotation are 390.63 kHz and -358.31 kHz.
%These are roughly consistent with the detuning.
}
\label{fig:storage_PhaseRotation}
\end{figure}
In the main text, we only highlight the result for storing and releasing the two-mode entangled state $(\alpha,\beta,\theta)=(1/\sqrt{2},1/\sqrt{2},0)$.
Here, in the following the results for various states  $(\alpha,\beta,\theta)=(1/\sqrt{2},1/\sqrt{2},0),(1/\sqrt{3},\sqrt{2/3}, \pi),\\ (1/\sqrt{2},1/\sqrt{2}, \pi)$, and $(1/\sqrt{2},1/\sqrt{2}, 5\pi/6)$ are shown as for different emission timings.

\subsection{Entangled state: $(\alpha,\beta,\theta)=(1/\sqrt{2},1/\sqrt{2},0)$}
Figure~\ref{fig:storage_DMDelay} shows the full results corresponding to Fig.~4 in the main text, storing $(\ket{0,1}+\ket{1,0})/\sqrt{2}$ with different emission timing for each mode, including the real and imaginary parts of the density matrices.
The imaginary parts of the off-diagonal elements $\ket{0,1}\bra{1,0}$ and $\ket{1,0}\bra{0,1}$ are increasing or decreasing in proportion to the storage time, similar to Ref. [16].
This is because the emission timing of each mode is different, even though the total photon number of the states are equal.
Figure \ref{fig:storage_PhaseRotation} shows the phase rotation of the quantum states caused by the detuning.
The phase corresponds to the angle of an off-diagonal component of the obtained density matrix $\rho_{\ket{1,0}\bra{0,1}}$.
The slope means the rotation frequency, and the frequencies are $390$ kHz and $-358$ kHz.
These are roughly consistent with the detuning.

\begin{figure*}
\includegraphics[width = 0.9\textwidth]{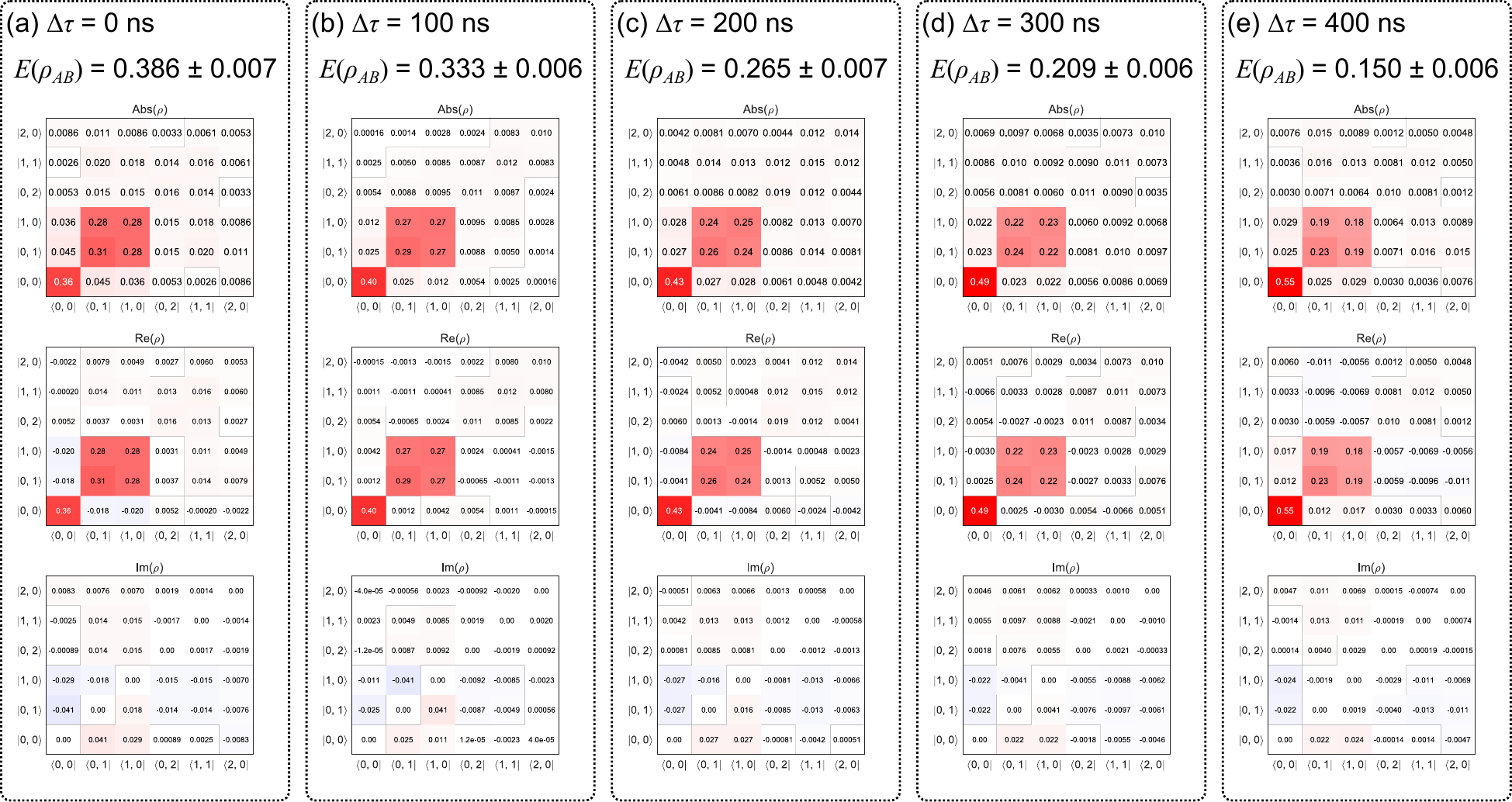}
\caption{Experimental results of storing $(\ket{0,1} + \ket{1,0})/\sqrt{2}$.
First row: absolute values of obtained density matrices.
Second row: real part of obtained density matrices.
Third row: imaginary part of obtained density matrices.
}
\label{fig:storage_DM}
\end{figure*}

In this experiment, we measured the simultaneous emission of each mode of $(\ket{0,1}+\ket{1,0})/\sqrt{2}$.
Since the emitted state is a state of fixed total photon number, the phase rotation does not occur under simultaneous emission.
Figure~\ref{fig:storage_DM} shows the full results of this experiment.
Unlike the experimental situation as explained in the main text, the imaginary part of the density matrix does not change in proportion to the storage time.
In the following, we show the results with other parameters for simultaneous emission timings, and where the phase rotation does not occur.

\begin{figure*}
\includegraphics[width = 0.9\textwidth]{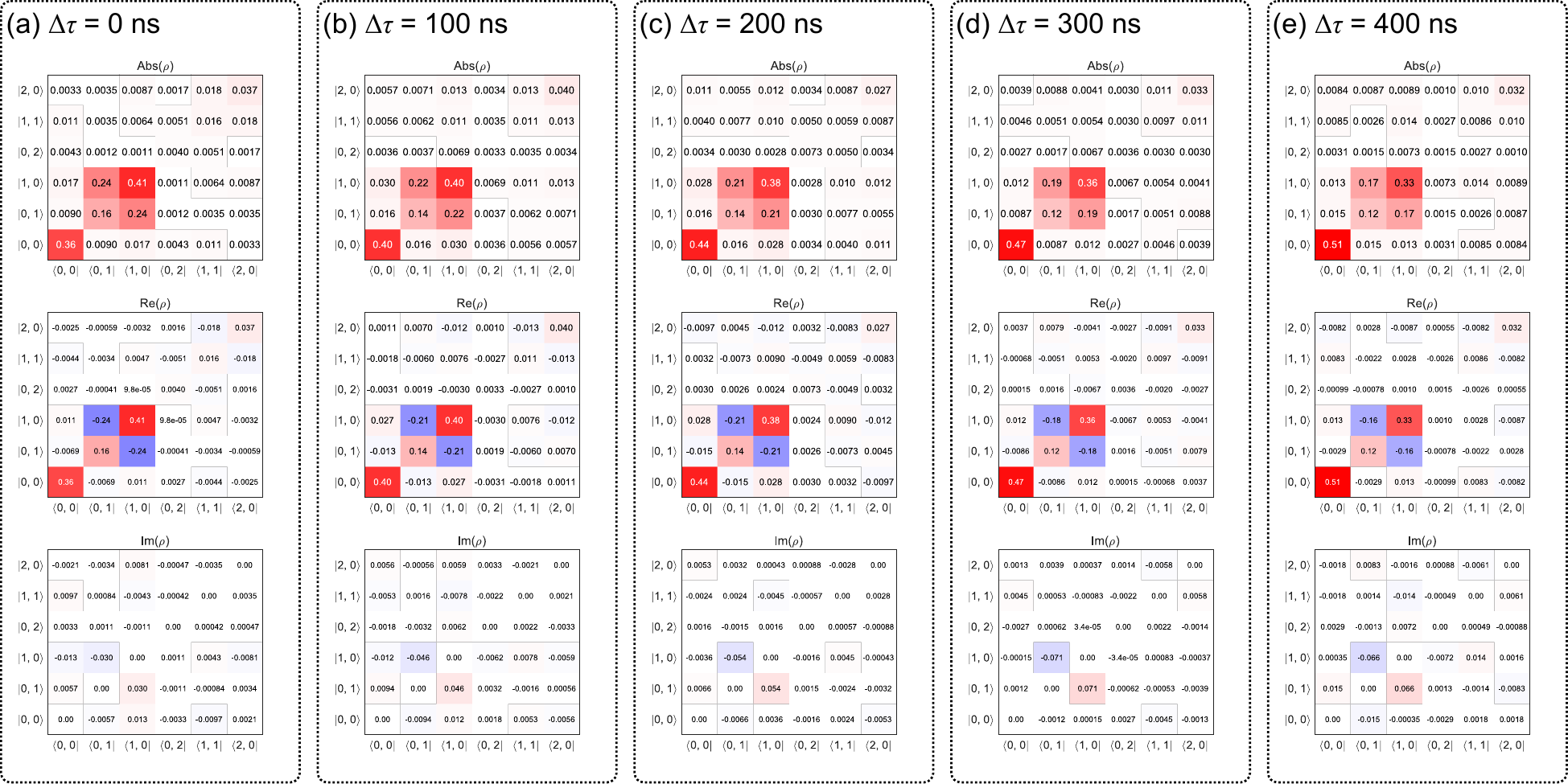}
\caption{Experimental results of storing $(\ket{0,1} +\text{e}^{\text{i}\pi}\sqrt{2}\ket{1,0})/\sqrt{3}$.
First row: absolute values of obtained density matrices.
Second row: real part of obtained density matrices.
Third row: imaginary part of obtained density matrices.
}
\label{fig:storage_DM2;1}
\end{figure*}
\begin{figure}
\includegraphics[scale = 0.8]{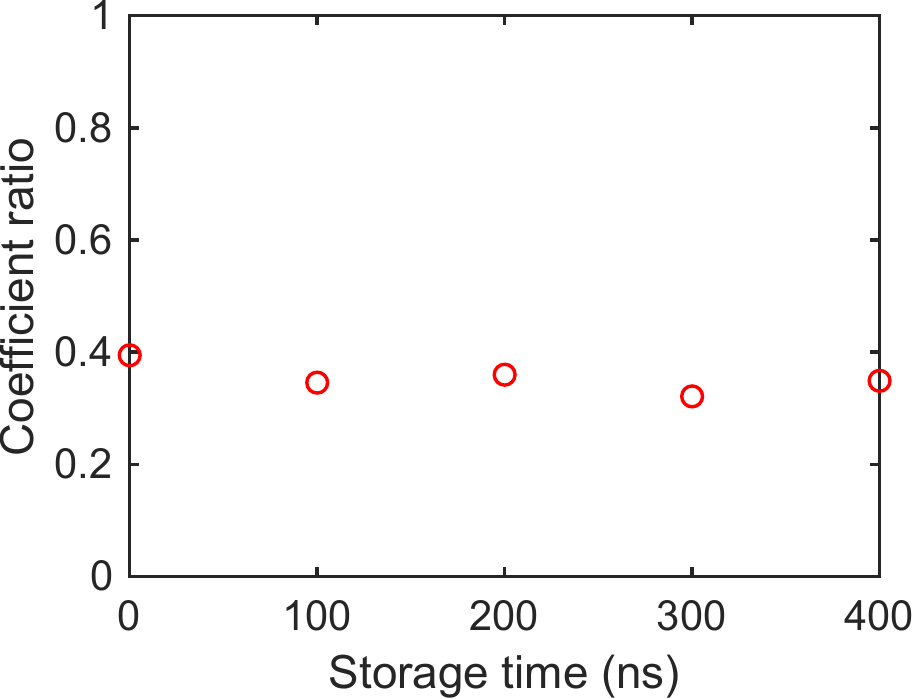}
\caption{Experimentally obtained coefficient ratio of $\ket{0,1}\bra{0,1}$ and $\ket{1,0}\bra{1,0}$ of $(\ket{0,1}+\text{e}^{\text{i}\pi}\sqrt{2}\ket{1,0})/\sqrt{3}$.
%Ratio: rho_{\ket{0,1}\bra{0,1}/rho_{\ket{1,0}\bra{1,0}}}
}
\label{fig:storage_coefficient}
\end{figure}

\subsection{Entangled state: $(\alpha,\beta,\theta)=(1/\sqrt{3},\sqrt{2/3}, \pi)$}
In Fig.~\ref{fig:storage_DM2;1}, the density matrices of $(\ket{0,1}+\text{e}^{\text{i}\pi}\sqrt{2}\ket{1,0})/\sqrt{3}$ are generated and stored with simultaneous emission.
To change the coefficients, only transmittance and reflectance of the beamsplitter combining the idler fields have to be changed in principle.
In this experiment, we replace the 50:50 BS by a 33:67 BS.
Diagonal terms which are components of $\ket{0,1}\bra{0,1}$ and $\ket{1,0}\bra{1,0}$ are different from each other.
In Fig.~\ref{fig:storage_coefficient}, the squared coefficient ratio $|\alpha/\beta|^2$ is kept regardless of the storage.
Since the experimentally obtained squared ratio is $\sim$0.4, the actual state is $\sim\ket{0,1}-1.22\ket{1,0}$.

\begin{figure}
\includegraphics[scale = 1]{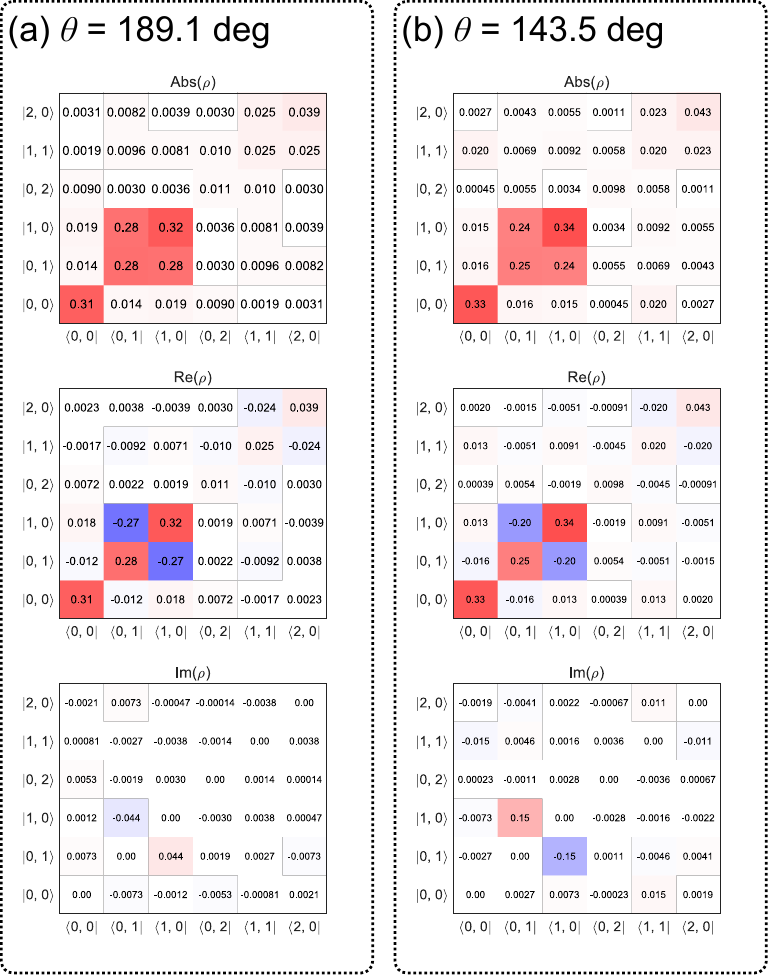}
\caption{Experimental results of $(\ket{0,1}+\text{e}^{\text{i}\pi}\ket{1,0})/\sqrt{2}$ (left side) and $(\ket{0,1} +\text{e}^{\text{i}5\pi/6} \ket{1,0})/\sqrt{2}$ (right side).
First row: absolute values of obtained density matrices.
Second row: real part of obtained density matrices.
Third row: imaginary part of obtained density matrices.
And the phase of each density matrix (the angle of the term $\ket{1,0}\bra{0,1}$) is written on the top of density matrices.
}
\label{fig:storage_DM0deg30deg}
\end{figure}
\begin{figure*}
\includegraphics[width = 0.9\textwidth]{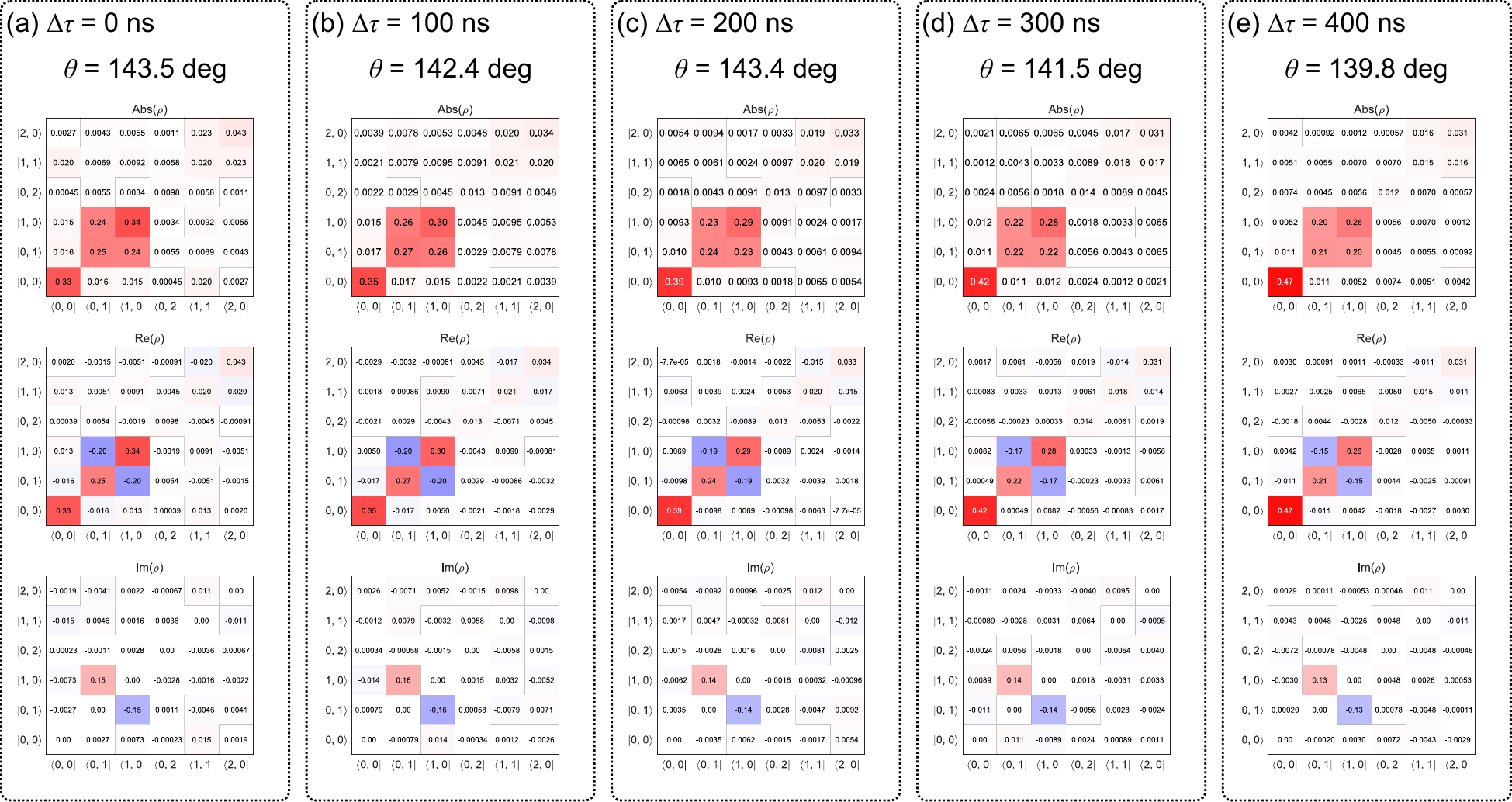}
\caption{Experimental results of storing $(\ket{0,1} +\text{e}^{\text{i}5\pi/6}\ket{1,0})/\sqrt{2}$.
First row: absolute values of obtained density matrices.
Second row: real part of obtained density matrices.
Third row: imaginary part of obtained density matrices.
And the phase of each density matrix (the angle of the term $\ket{1,0}\bra{1,0}$) is written on the top of density matrices.
}
\label{fig:storage_DM30deg}
\end{figure*}

\subsection{Entangled states: $(\alpha,\beta,\theta)=(1/\sqrt{2},\sqrt{2}, \pi)$ and $(1/\sqrt{2},\sqrt{2},5\pi/6)$}

Moreover, we generated other types of states: $(\ket{0,1}+\text{e}^{\text{i}\pi}\ket{1,0})/\sqrt{2}$ and $\ket{0,1}+\text{e}^{\text{i}5\pi/6}\ket{1,0}$.
For this purpose, to change the phase, the lock phase between idler probes at BS combining the idler fields has to be changed.
The phase was changed by shifting the offset of the error signal of the interferometer.
The comparison is shown in Fig.~\ref{fig:storage_DM0deg30deg}.
We can see the difference between $(\ket{0,1}+\text{e}^{\text{i}\pi}\ket{1,0})/\sqrt{2}$ and $(\ket{0,1}+\text{e}^{\text{i}5\pi/6}\ket{1,0})/\sqrt{2}$, especially in the imaginary part of the off-diagonal term $\ket{0,1}\bra{1,0}$, where the storage time is set to zero.

In Fig.~\ref{fig:storage_DM30deg} the density matrices of $(\ket{0,1}+\text{e}^{\text{i}5\pi/6}\ket{1,0})/\sqrt{2}$ are shown for different emission timings.
The phases are kept during storage since the imaginary parts are not changed.

%%%%%%%%%%%%%%%%
\section{Logarithmic negativity}
Here we discuss the entanglement of the released states.
Because the states are suporpositions of $\ket{1,0}$ and $\ket{0,1}$, they should become two-mode entangled states.
Therefore, the logarithmic negativity, which at the least can distinguish whether entanglement is present or not, can be useful to verify the released states.
We calculate the logarithmic negativity of the released states only in the subspace of $\{\ket{0,0}, \ket{1,0}, \ket{0,1}, \ket{1,1}\}$.
%Since considering this subspace is just limiting the upper photon number of each memory, then this operation is local operation, and it does not make new entanglement.
%Therefore, we can say that if there exists entanglement in this subspace, there exists entanglement in full space.
By only considering that subspace we will never add entanglement on top of what is really present in the state (we may just reduce it).
Those local projections or truncations are non-unitary and typically decrease the entanglement.
In other words, if we obtain a non-zero logarithmic negativity in the subspace this will prove that the the amount of entanglement in the full space is at least as large as given by the obtained value.
As a result, any non-zero value for the logarithmic negativity gives a lower bound of the entanglement in the full space.

A well-known fact is that any entangled state has non-zero value of logarithmic negativity. 
On the other hand, we cannot say that there does not exist entanglement at all if the state has zero logarithmic negativity.
Non-zero logarithmic negativity is apparently a sufficient condition for entanglement, but not a necessary condition. 
Although it is generally a difficult task to derive a necessary and sufficient condition for entanglement, fortunately there are various known sufficient conditions for entanglement. 
In particular, we employ a criterion based on the eigenvalues of the partially transposed density matrix.
If a density matrix $\hat{\rho}_{AB}$ for two harmonic oscillators A and B represents a statistical mixture of Gaussian states, the following $E(\hat{\rho}_{AB})$ is always nonnegative:
\begin{align}
E(\hat{\rho}_{AB}) = \log||\hat{\rho}_{AB}^{T_B}||
\end{align}
where $A, B$ denotes the two subsystems, and $\cdot^{T_B}$ means the transposition for subsystem $B$.
$||\cdot||$ means trace norm, which corresponds to the sum of absolute eigenvalues.
Partial transposition for subsystem $B$ means time-reversal only for $B$.
If the state is not entangled, since the time-reversed state of $B$ is still an physical state, the eigenvalues of the partial transposition of the density matrix is nonnegative.
Then the sum of absolute eigenvalues becomes 1, and the logarithmic negativity $E(\hat{\rho}_{AB})$ becomes 0.
Therefore if the logarithmic negativity is nonzero, we can say that the state is entangled.

Experimentally obtained logarithmic negativities $E(\hat{\rho})$ for the storage 0 ns, 100 ns, 200 ns, 300 ns, 400 ns of $(\ket{0,1}+\ket{1,0})/\sqrt{2}$ are $0.386\pm0.007, 0.333\pm0.006, 0.265\pm0.007, 0.209\pm0.006, 0.150\pm0.006$, respectively. 
We can see that every logarithmic negativity has positive values within errors, where the errors are calculated by the bootstrap method.

%%%%%%%%%%%%%%%%
\section{Phase-fluctuation analysis}
From the experimental results, we have seen that the phase-preserving feature of our memory system is in the quantum regime, enough to preserve the off-diagonal elements of a density matrix.
Here, we make an estimation of the amount of phase fluctuations in our system from the reconstructed density matrices. 

\begin{figure}[tb]
\centering
\includegraphics[scale = 0.8]{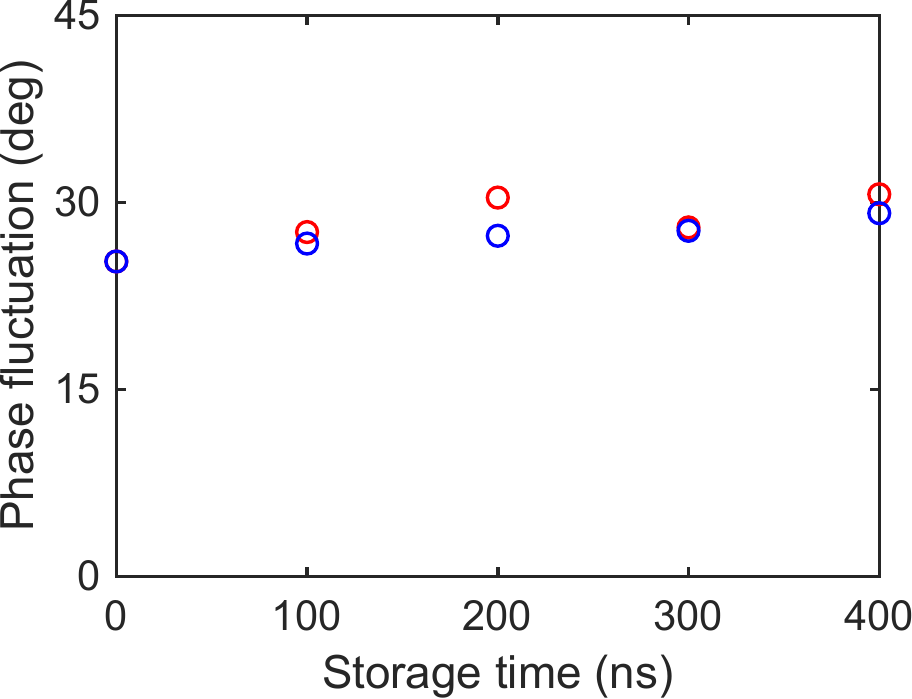}
\caption{
Standard deviations of phase fluctuations $\sigma$ estimated from experimental density matrices for entangled state $(\ket{0,1}+\ket{1,0})/\sqrt{2}$ and various storage times. 
}
\label{fig:dephasing}
\end{figure}

The reconstructed density matrices deviate from the ideal ones. 
Here we separate the deviations into optical losses (or amplitude damping) and phase fluctuations (or dephasing). 
We consider a density matrix $\hat{\rho}=\sum_{k,\ell,m,n\in\{0,1\}}\rho_{k,\ell,m,n}\ket{k,\ell}\bra{m,n}$ in the number basis up to one photon for each side.

Optical losses of memories can be estimated by the experimental data of single photon storage.
Fake counts affect the density matrix of the experimentally obtained state, and this time fake counts are about 10 cps.
We can estimate the phase fluctuation by comparing the experimentally obtained state and the state as changed by losses from the ideal state.
From the density matrix of the ideal state, the fake counts affect the density matrix as follows:
\begin{align}
\hat{\rho} \rightarrow (1-L_\text{fake})\hat{\rho}+L_\text{fake}\ket{0,0}\bra{0,0}
\end{align}
where $L_\text{fake}$ denotes the loss by fake counts.
Then, the density matrix reduced by the optical loss becomes
\begin{align}
&\rho_{\ket{0,0}\bra{0,0}} \rightarrow \rho_{\ket{0,0}\bra{0,0}}+L_1\rho_{\ket{0,1}\bra{0,1}}+L_2\rho_{\ket{1,0}\bra{1,0}}\\
&\rho_{\ket{0,1}\bra{0,1}} \rightarrow (1-L_1)\rho_{\ket{0,1}\bra{0,1}}\label{eq:alpha}\\
&\rho_{\ket{1,0}\bra{1,0}} \rightarrow (1-L_2)\rho_{\ket{1,0}\bra{1,0}}\label{eq:beta}\\
&\rho_{\ket{0,1}\bra{1,0}} \rightarrow \sqrt{(1-L_1)(1-L_2)}\rho_{\ket{0,1}\bra{1,0}}\\
&\rho_{\ket{1,0}\bra{0,1}} \rightarrow \sqrt{(1-L_1)(1-L_2)}\rho_{\ket{1,0}\bra{0,1}}
\end{align}
where $L_1, L_2$ denote losses of CCS-1 and CCS-2, respectively.
%The remaining difference between \red{the} experimentally obtained state and state changed by losses from ideal state is off-diagonal term.
Next, let us consider the phase fluctuation effect.
Assuming a phase fluctuation in a Gaussian distribution $(1/\sqrt{2\pi\sigma^2})\text{e}^{-\theta^2/(2\sigma^2)}$ with a standard deviation $\sigma$, the phase fluctuation changes the density matrix as
\begin{align}
&\rho_{\ket{0,1}\bra{1,0}} \rightarrow \text{e}^{-\frac{\sigma^2}{2}}\rho_{\ket{0,1}\bra{1,0}}\\
&\rho_{\ket{1,0}\bra{0,1}} \rightarrow \text{e}^{-\frac{\sigma^2}{2}}\rho_{\ket{1,0}\bra{0,1}}
\end{align}
This is derived from the integral $\int \text{e}^{\text{i}\theta}\text{e}^{-\theta^2/(2\sigma^2)}\text{d}\theta = \int \text{e}^{-(\theta-i\sigma^2)^2/(2\sigma^2)}\text{e}^{-\sigma^2/2}\text{d}\theta = \sqrt{2\pi\sigma^2}\text{e}^{-\sigma^2/2}$. 
Therefore, by comparing the off-diagonal terms, we can estimate $\sigma$, i.e. the phase fluctuation.
Moreover, taking into account multiphoton contributions further complicates the situation. 
However, here we neglect such imperfections and simplify the situation by just considering the two factors, and taking the subspace of the density matrix up to one photon for each side.

Our assumption in this analysis is that the initial quantum states are pure entangled states $\alpha\ket{0,1}+\beta \text{e}^{\text{i}\theta}\ket{1,0}$, where $\alpha$ and $\beta$ are estimated by comparing the diagonal terms of the experimentally obtained density matrix with those of the ideal states with losses through (\ref{eq:alpha}) and (\ref{eq:beta}).
Then, $\alpha$ and $\beta$ are estimated 0.71 and 0.64, respectively while $1/\sqrt{2} \simeq 0.71$.
We can tune $\alpha$, $\beta$ by tuning the pump power for each memory.
Then, we calculate the amounts of losses $L_1, L_2$ by the single photon storage experiment.
Moreover, we calculate the phase fluctuations $\sigma$ consistent with the experimental density matrix $\hat{\rho}$, where the subspace of up to one photon for each side is taken and renormalized (the neglected multiphoton components are about 3\%). 
The off-diagonal terms, together with the obtained losses, gives information on the phase fluctuations $\sigma$ by the relation
\begin{align}
\text{e}^{-\frac{\sigma^2}{2}}=\frac{\lvert\rho_{\ket{0,1}\bra{1,0}}^\text{exp}\rvert}{\sqrt{\lvert\rho_{\ket{0,1}\bra{0,1}}^\text{exp}\rvert\lvert\rho_{\ket{1,0}\bra{1,0}}^\text{exp}\rvert}}.
\end{align}

The calculated photon-preserving efficiencies $1-L_1, 1-L_2$ and phase fluctuations $\sigma$ are shown in Fig.~\ref{fig:singlephoton} and Fig.~\ref{fig:dephasing}, respectively.
As for the phase fluctuations in Fig.~\ref{fig:dephasing}, the results are from 25$^\circ$ to 30$^\circ$, but we could not find a clear tendency like a worsening dephasing during storage. 
It seems that other accidental factors obscure the accumulating dephasing during storage. 
The calculated phase fluctuations may look a bit large, but we note that amplitude fluctuations may be converted to phase fluctuations in this analysis. 
If the effective pump power fluctuates, the single-photon fraction of the heralded state fluctuates, making a statistical mixture which has smaller off-diagonal elements, resulting in larger $\sigma$.
We are aware that fluctuations in the heralding event rate during experiments are much larger than those expected from the Poisson distribution.
In that sense, the calculated values are considered as upper limits of the actual phase fluctuations in our system. 
Moreover, even for the actual phase fluctuations, we are not sure about whether the phase-instability is caused by the memory cavity itself or by the laser system supplying the LO, the pump, and the displacement beam.

\end{document}